\documentclass[journal]{IEEEtran}
\usepackage{graphicx}
\usepackage{multirow,multicol}
\usepackage{amsmath}
\usepackage{setspace}
\usepackage{booktabs}
\usepackage{cellspace}%
\usepackage{makecell}
\usepackage{amsfonts}
\usepackage{url}
\usepackage[caption=false,font=footnotesize]{subfig}
\usepackage{cite}
\usepackage{color, colortbl}
\definecolor{Gray}{RGB}{255, 248, 240}

\usepackage{pifont}

\begin{document}

\title{Asymmetric Encoder-Decoder Based on Time-Frequency Correlation for Speech Separation}
\author{Ui-Hyeop Shin and Hyung-Min Park, \IEEEmembership{Senior Member, IEEE}
\thanks{This work was supported in part by Institute of Information \& communications Technology Planning \& Evaluation (IITP) grant funded by the Korea government(MSIT)(RS-2022-II220989, Development of Artificial Intelligence Technology for Multi-speaker Dialog Modeling), and in part by Institute of Information \& communications Technology Planning \& Evaluation (IITP) grant funded by the Korea government(MSIT) (RS-2022-II220621, Development of artificial intelligence technology that provides dialog-based multi-modal explainability).}
\thanks{The authors are with the Department of Electronic Engineering, Sogang University, Seoul 04107, Republic of Korea (e-mail: hpark@sogang.ac.kr).}}

\markboth{Journal of \LaTeX\ Class Files,~Vol.~14, No.~8, August~2015}%
{Shell \MakeLowercase{\textit{et al.}}: Bare Demo of IEEEtran.cls for IEEE Journals}

\maketitle

\begin{abstract}
Speech separation in realistic acoustic environments remains challenging because overlapping speakers, background noise, and reverberation must be resolved simultaneously. Although recent time--frequency (TF) domain models have shown strong performance, most still rely on late-split architectures, where speaker disentanglement is deferred to the final stage, which can limit speaker-discriminative modeling under adverse conditions. To address this issue, we propose SR-CorrNet, an asymmetric encoder--decoder framework that introduces the separation-reconstruction (SepRe) strategy into a TF dual-path backbone. The encoder performs coarse separation from mixture observations, while the weight-shared decoder progressively reconstructs speaker-discriminative features with cross-speaker interaction, enabling stage-wise refinement. To complement this architecture, we formulate speech separation as a structured correlation-to-filter problem: spatio-spectro-temporal correlations computed from the observations are used as input features, and the corresponding deep filters are estimated to recover target signals. We further incorporate an attractor-based dynamic split module to adapt the number of output streams to the actual speaker configuration. Experimental results on WSJ0-\{2,3,4,5\}mix, WHAMR!, LibriCSS, and AMI demonstrate strong separation and recognition performance across anechoic, noisy-reverberant, and real-recorded conditions in both single- and multi-channel settings, highlighting the effectiveness of TF-domain SepRe with correlation-based filter estimation for speech separation.
\end{abstract}

\begin{IEEEkeywords}
Correlation, multi-channel source separation, continuous speech separation, Transformer, LibriCSS
\end{IEEEkeywords}

\IEEEpeerreviewmaketitle

\section{Introduction}

\IEEEPARstart{S}{peech} separation is a fundamental task in speech and audio signal processing, serving as a key front-end for applications such as meeting transcription, hands-free communication, hearing aids, and distant automatic speech recognition (ASR). 
In realistic acoustic environments, speech recordings often contain multiple overlapping speakers, background noises, and reverberant reflections, making robust separation a persistent challenge.
Time-domain approaches first drove rapid progress, beginning with TasNet~\cite{luo_tasnet_2018, luo_conv-tasnet_2019} and evolving through DPRNN~\cite{luo_dual-path_2020} and Sepformer~\cite{subakan_attention_2021}. 
Despite their success, these models operate on short analysis windows and must implicitly capture long-range convolutional effects, which limits robustness in reverberant and noisy conditions. 
To address this, recent studies have increasingly adopted dual-path modeling of time-frequency (TF) representations in the short-time Fourier transform (STFT) domain~\cite{yang_tfpsnet_2022, wang_tf-gridnet_2023_ICASSP, saijo_tf-locoformer_2024, li_spmamba_2024}, where networks operate on complex spectra and estimate either complex masks or direct spectrogram mappings. 
Owing to its explicit spectro-temporal structure and strong representation capability, this formulation has been widely used not only for separation but also for speech enhancement tasks including dereverberation and denoising~\cite{dang_dpt-fsnet_2022, abdulatif_cmgan_2024, lu_mp-senet_2023, chao_investigation_2024}.

In multi-channel scenarios, early systems incorporated spatial cues such as the inter-channel phase difference (IPD) by concatenating them with spectral features (e.g.~\cite{bahmaninezhad_comprehensive_2019,chen_multi-channel_2018,chen_continuous_2021,yoshioka_multi-microphone_2018,wang_combining_2019,gu_multi-modal_2020, yoshioka_vararray_2022}). 
However, treating spectral and spatial streams independently yielded limited performance gains. 
Extensions of TasNet-style models to multi-channel inputs~\cite{luo_fasnet_2019, chen_beam-guided_2022 ,pandey_tparn_2022} faced similar challenges, as the network was still required to implicitly learn spatial coherence from unstructured observations.
Neural beamforming methods partially addressed this by estimating spatial filters in a data-driven manner~\cite{zhang_adl-mvdr_2021, zhang_all-neural_2022, xu_generalized_2021, li_mimo_2021}, yet most recent multi-channel frameworks continue to stack real and imaginary components of multi-channel complex spectra and rely on direct complex mapping, effectively delegating spatial reasoning to the network itself~\cite{wang_count_2021, tan_neural_2022, wang_tf-gridnet_2023, quan_spatialnet_2024}.

Filter-based formulations have also been studied, such as deep multi-frame minimum variance distortionless response (MVDR) filtering that computes the filter from network-estimated correlation statistics~\cite{tammen_deep_2021} and deep non-linear multi-channel filters~\cite{tesch_insights_2023}, where the filters or their statistics are estimated from the spectra.
For TF-domain separation, cross-band and narrow-band processing has been exploited~\cite{kalkhorani_tf-crossnet_2024, quan_narrow-band_2022}, and unified models have been developed for diverse input conditions and tasks including speaker counting~\cite{zhang_toward_2023, saijo_single_2023}, and for conversational speech with neural diarization~\cite{taherian_conversational_2024} or attractor-based separation of an unknown number of speakers~\cite{wang_attractor_2025}.
Across both single- and multi-channel settings, the field has largely converged on TF-domain complex spectral dual-path processing as the dominant paradigm.

\begin{figure}
\centering
\subfloat[Late-split]{\includegraphics[width=0.90\columnwidth]{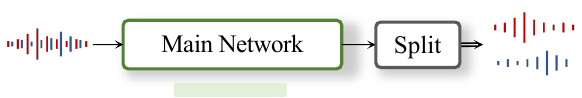}} \\[.5pt]
\subfloat[Early-split]{\includegraphics[width=0.95\columnwidth]{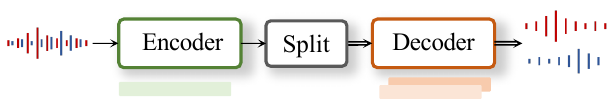}}
\footnotesize
\caption{{Block diagrams of (a) late-split  and (b) early-split schemes.}}
\label{fig:split_scheme}
\vspace{-3mm}
\end{figure}

Despite this progress, most existing TF-domain systems still depend heavily on implicit learning in two major aspects. 
First,  as illustrated in Fig.~\ref{fig:split_scheme}(a), the predominant architectures adopt a late-split design, where mixture features are compressed into a single latent representation and speaker separation occurs only at the final output stage. 
In this design, the shared representation must embed all speaker information together until the final stage, which can weaken discriminability and lead to speaker confusion, particularly under reverberant and noisy conditions where overlapping spectral structures are harder to resolve.
Second, the input–output formulation itself lacks explicit structure: networks typically receive raw complex STFT coefficients, or their concatenation across channels, and directly predict clean spectra or time-frequency masks.
This direct mapping paradigm requires the model to implicitly discover and exploit the physical dependencies that govern the observed signals, such as inter-channel spatial coherence and the temporal continuity imposed by reverberation, rather than encoding them as part of the processing pipeline, although such structure can be provided explicitly as a useful inductive bias. 

These two aspects—late-split architectures and unstructured input–output formulations—motivate the two complementary directions pursued in this work.
First, we reformulate TF-domain separation as an explicit separation-reconstruction process, where the encoder performs early speaker disentanglement and the decoder progressively reconstructs speaker-discriminative features through shared and cross-interacting modules. 
Second, we replace the conventional direct-mapping formulation with a structured correlation-to-filter scheme, in which spatio-spectro-temporal correlations computed from the observed signals serve as input features for estimating the corresponding separation filters.

\begin{figure}
\footnotesize
\centering
\includegraphics[width=0.995\columnwidth]{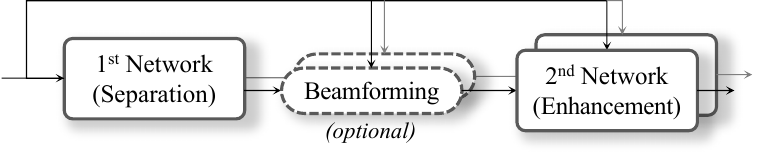}
\caption{{Illustration of two-stage multi-channel separation structure.}}
\vspace{-2mm}
\label{fig:2stage}
\end{figure}

\textbf{Previous Works} 
Our earlier work~\cite{shin_separate_2024} proposed the Separation-Reconstruction (SepRe) strategy as an alternative to the late-split design.
As illustrated in Fig.~\ref{fig:split_scheme}(b), an asymmetric encoder-decoder scheme performs early separation from the mixture in the encoder, and progressively reconstructs speaker-discriminative features in the decoder through weight-shared and cross-interacting modules.
This early-split, stage-wise reconstruction was shown to enhance speaker discrimination and improve efficiency.
This model architecture also exhibits similarities to the widely adopted two-stage pipeline in multi-channel systems~\cite{wang_multi-microphone_2020, wang_multi-microphone_2021, pandey_multichannel_2022, wang_tf-gridnet_2023, shin_tf-corrnet_2025}. 
As shown in Fig.~\ref{fig:2stage}, conventional approaches typically cascade a first-stage separation network with an optional beamforming step and a second-stage enhancement network, where each stage addresses a complementary subtask.
The SepRe framework can be regarded as an end-to-end alternative that internalizes this two-stage structure within a single model, enabling direct feature-level interaction between coarse separation and fine-grained reconstruction. 
However, the original SepReformer was developed in the time domain, inheriting the limitations of short analysis windows and implicit spectro-temporal modeling.
Adapting this framework to the TF domain is not straightforward, as it must accommodate complex spectral processing, multi-channel spatial dependencies, and filter-based output formulations that do not arise in time-domain models.

On the representational side, our previous work introduced a correlation-based framework for multi-channel separation, termed TF-CorrNet~\cite{shin_tf-corrnet_2025}. 
Unlike conventional methods that concatenate magnitude and IPD or stack real and imaginary components across channels, TF-CorrNet directly computes inter-channel correlations between microphone signals to capture spatial information. 
A generalized phase transform (PHAT-$\beta$)~\cite{donohue_performance_2007} is applied to balance spectral and spatial emphasis, and the resulting correlation features are processed within a TF dual-path backbone. 
Our concurrent preprint, IF-CorrNet~\cite{shin_deep_2026}, extended the correlation-to-filter paradigm along the temporal dimension for single-channel dereverberation of a single speaker. In contrast, this work addresses multi-speaker separation by jointly exploiting spatial, temporal, and cross-band spectral correlations with the corresponding filters for both single- and multi-channel inputs.

\textbf{Proposed Method} 
Building upon these works, this paper presents \textit{SR-CorrNet} (Separation and Reconstruction of Correlation), a unified framework that combines the SepRe strategy with the correlation-to-filter paradigm within a TF dual-path backbone. 
The encoder analyzes spatio-spectro-temporal correlations computed from multi-channel observations to perform coarse separation, and the decoder progressively reconstructs speaker-discriminative features through weight-shared modules with cross-speaker interaction. An attractor-based dynamic split module further adapts the number of output streams to the actual speaker configuration. 
Extensive experiments on WSJ0-\{2,3,4,5\}mix, WHAMR!, LibriCSS, and the AMI meeting corpus demonstrate that SR-CorrNet achieves strong performance across anechoic, noisy-reverberant, and real-recorded conditions under both single-channel and multi-channel settings.

\vspace{0.5mm}
\noindent
The contributions of this work are summarized as follows:
\begin{itemize}
\item 
We propose a unified deep filtering model where spatio-spectro-temporal correlations are provided as input features, motivated as an inductive bias for filter estimation, and the complex filters with the corresponding extent are estimated as outputs. It handles single- and multi-channel, noisy, and reverberant conditions, and is extended to multiple reference channels for robustness.
\vspace{1mm}
\item 
We adapt the SepRe strategy~\cite{shin_separate_2024} to the TF-domain correlation-to-filter framework, where an asymmetric encoder-decoder performs early separation and progressive reconstruction with weight-shared and cross-speaker modules and stage-wise supervision.
\vspace{1mm}
\item 
We integrate attractor-based speaker counting and conditioning~\cite{lee_boosting_2024} into the separator, adapting the number of output streams without an external counting network and skipping the decoder for non-speech chunks.
\end{itemize}

\section{Correlation for Filter Estimation}
\subsection{Formulation}

Let the observation of $K$ multi-speakers from multi-channel microphones be given as $\mathbf{x}_{tf} = [\hspace{-.2mm}{{X}_{\hspace{-.2mm}tf1\hspace{-.1mm}},...,\hspace{-.2mm}{X}_{tfM}}\hspace{-.2mm}]^T \hspace{-.5mm}\in\hspace{-.2mm}\mathbb{C}^{M}\hspace{-.3mm},1\hspace{-.2mm}\le\hspace{-.2mm}t\hspace{-.2mm}\le\hspace{-.2mm}T, 1\hspace{-.5mm}\le\hspace{-.5mm} f\hspace{-.5mm}\le\hspace{-.5mm}F$ where $T$ and $F$ are the number of time frames and frequency bins, respectively. Then, we can formulate the input observation as
\begin{equation}
    \mathbf{x}_{tf} = \sum_{k=1}^K\left(\mathbf{h}_{k,f} S_{k,tf} + \sum_{\tau=0}^{L_0} \mathbf{r}_{k,\tau f} S_{k,(t-\tau)f}\right) + \mathbf{n}_{tf},
    \label{eq:input}
\end{equation}
where ${S}_{k,tf}$ and $\mathbf{h}_{k,f}\hspace{-1mm}=\hspace{-1mm}[H_{k,f1},...,H_{k,fM}]\in\hspace{-.5mm}\mathbb{C}^{M}\hspace{-.6mm}$ denote a speech source and the corresponding direct-path relative transfer function of the $k$-th speaker. On the other hand, $\mathbf{r}_{k,\tau f}\in\mathbb{C}^{M}$ is the convolutive transfer function from early reflections and late reverberations up to $L_0$ frames. $\mathbf{n}_{tf}\hspace{-1mm}\in\hspace{-.5mm}\mathbb{C}^{M}\hspace{-.6mm}$ is an additive noise. The goal is to separate and obtain the desired direct signals $S_{k, tf1}\hspace{-.3mm}=\hspace{-.3mm}H_{k,f1} S_{k, tf}\hspace{-.5mm}\in\hspace{-.5mm}\mathbb{C}$ from $\mathbf{x}_{tf}$ by separating the speech signals and suppressing the early reflection, late reverberation, and additive noise\footnote{Assuming a multi-input single-output (MISO) model when the first channel $m=1$ is the reference channel for simplicity.}.

\subsection{Spatio-spectro-temporal correlation as an input feature}

To explicitly extract spatial information from the multi-channel observation $\mathbf{x}_{tf}$, we compute spatial correlations as
\begin{equation}
    Z_{tfm_1m_2} = {X}_{tfm_1}{X}^*_{tfm_2},
\end{equation}
where $1\le m_1, m_2 \le M$,
or equivalently, in a matrix form, represented as
\begin{equation}
\mathbf{Z}_{tf}=\mathbf{x}_{tf}\mathbf{x}^H_{tf} \in \mathbb{C}^{M\times M}.
\end{equation}
Then, these input features naturally include power information on diagonal elements while IPDs are included as off-diagonal elements.

On the other hand, the convolutive transfer function terms \(\mathbf{r}_{k, \tau f}\) induce strong correlations between the current frame \(\mathbf{x}_{tf}\) and both its past and future frames. 
By explicitly computing these inter-frame correlations, the model receives a direct cue of the reverberant structure imposed by \(\mathbf{r}_{k, \tau f}\) for the multi-tap deep filter estimation. 
Furthermore, the formulation of (\ref{eq:input}) can be represented more accurately by considering the convolution effects over frequencies from a finite window length of the STFT as
\begin{equation}
    \mathbf{x}_{tf} = \sum_{k=1}^K\sum_{f'=1}^{F} \left( \mathbf{h}^{(f')}_{k,f} S_{k,tf'} \\ + \sum_{\tau=0}^{L_0} \mathbf{r}^{(f')}_{k,\tau f} S_{k,(t-\tau)f'} \right) + \mathbf{n}_{tf},
    \label{eq:input2}
\end{equation}
where $\mathbf{h}^{(f')}_{k,f}$ and $\mathbf{r}^{(f')}_{k,\tau f}$ are sets of band-to-band ($f'=f$) and cross-band ($f'\ne f$) components of the corresponding transfer functions, respectively. In particular, most of the energy is concentrated within adjacent frequencies range $f'\in [f-I_0, f+I_0]$, where $I_0$ denotes the number of adjacent frequency bins. 
Eq. (\ref{eq:input}) can be seen as an approximation by regarding $\mathbf{h}^{(f')}_{k,f}=0$ and $\mathbf{r}^{(f')}_{k,\tau f}=0$ for $f'\ne f$ assuming $I_0=0$.

Therefore, considering (\ref{eq:input2}), the input component $\mathbf{x}_{tf}$ is correlated not only across adjacent time frames but also over neighboring frequency bins. 
This allows the instantaneous spatial correlation to be extended into a generalized \textit{spatio-spectro-temporal} correlation, formulated as
\begin{equation}
\mathbf{Z}_{tf} = \tilde{\mathbf{x}}_{tf}\tilde{\mathbf{x}}^H_{tf} \in \mathbb{C}^{M(2L+1)(2I+1)\times M(2L+1)(2I+1)},
\label{eq:corr_MIMO}
\end{equation}
where
\begin{align}
\tilde{\mathbf{x}}_{tf}\hspace{-.5mm} &= [\bar{\mathbf{x}}^T_{t(f-I)}, ..., \bar{\mathbf{x}}^T_{tf}, ..., \bar{\mathbf{x}}^T_{t(f+I)}]^T \hspace{-.5mm}\in \mathbb{C}^{M(2L+1)(2I+1)},\\
\bar{\mathbf{x}}_{tf}\hspace{-.5mm} &= [\mathbf{x}^T_{(t-L)f}, ..., \mathbf{x}^T_{tf}, ..., \mathbf{x}^T_{(t+L)f}]^T \hspace{-.5mm}\in \mathbb{C}^{M(2L+1)}
\end{align}
represent multi-band and multi-frame vectors of multi-channel observations, where $2L+1$ and $2I+1$ are the numbers of adjacent frames and frequency bins of $\tilde{\mathbf{x}}_{tf}$, respectively. 
The vector $\tilde{\mathbf{x}}_{tf}$ can be conveniently obtained using the \textit{unfold} operation~\cite{paszke_pytorch_2019}.
However, the number of correlation components in $\mathbf{Z}_{tf}$ becomes extremely large. 
Assuming that the model is formulated under a MISO configuration and local correlations across time–frequency bins are sufficiently captured within the convolutional receptive field, 
the correlation can be computed with respect to a single reference observation (e.g., $m = 1$), while omitting redundant dependencies across channels, frames, and frequency bins, as
\begin{equation}
\mathbf{z}_{tf} = X_{tf1}\tilde{\mathbf{x}}^H_{tf} \in \mathbb{C}^{M(2L+1)(2I+1)}.
\label{eq:corr_MISO}
\end{equation}
Ideally, a 2d convolution (Conv2D) with a kernel size of $(2L{+}1)\times(2I{+}1)$ can aggregate such local correlations across neighboring time–frequency regions. 

Because (\ref{eq:corr_MISO}) relies on a single reference microphone, the correlation features can be sensitive to degradation of the reference channel, such as microphone occlusion or local noise. To alleviate this, all the channels at the current TF bin can be regarded as references:
\begin{equation}
\mathbf{z}_{tf} = \mathbf{x}_{tf}\tilde{\mathbf{x}}^H_{tf} \in \mathbb{C}^{M\times M(2L+1)(2I+1)},
\label{eq:corr_multiref}
\end{equation}
which consists of the rows of (\ref{eq:corr_MIMO}) associated with the current TF bin, at the cost of an $M$-fold increase in the input dimension. We use (\ref{eq:corr_MISO}) by default.

\begin{figure*}
\footnotesize
\centering
\includegraphics[width=2.03\columnwidth]{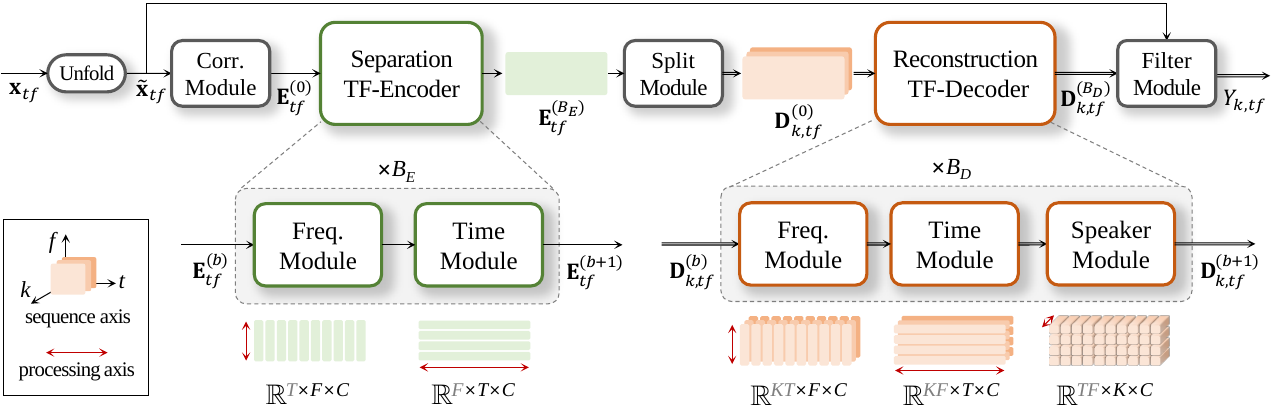}
\vspace{1mm}
\caption{{
Architecture of SR-CorrNet. From the multi-band and multi-frame vector of the multi-channel observation $\tilde{\mathbf{x}}_{tf}\in \mathbb{C}^{M(2L+1)(2I+1)}$, the correlation module computes the input feature $\mathbf{E}_{tf}^{(0)}\in\mathbb{R}^{C}$ of the TF-Encoder. The encoder output $\mathbf{E}_{tf}^{(B_E)}$ is split into speaker-specific features $\mathbf{D}_{k,tf}^{(0)}$, which are reconstructed by the TF-Decoder to predict the multi-channel multi-tap filter $\mathbf{w}_{k,tf}$ for $Y_{k,tf}$.
}}
\label{fig:SR_CorrNet}
\vspace{-1mm}
\end{figure*}

On the other hand, the raw correlations include large power-scale variations. To reduce them, the STFT coefficients of the variance-normalized waveform at each frequency are divided by their root-mean-square (RMS) value over time (Section~\ref{sec:setup_whamr} for WHAMR!) before computing the instantaneous correlations, which compensates these variations across frequencies. In addition, we adopt PHAT-$\beta$ weighting~\cite{shin_tf-corrnet_2025}, defined as
\begin{equation}
    Z_{tf,l,i,m} \gets 
    \frac{{X}_{tf1}{X}^*_{(t+l)(f+i)m}}
    {(|{X}_{tf1}{X}^*_{(t+l)(f+i)m}|{}+\varepsilon)^\beta},
\label{eq:PHAT}
\end{equation}
where $-L\hspace{-.5mm}\le l\le\hspace{-.5mm} L$, $-I\hspace{-.5mm} \le i \le\hspace{-.5mm} I$, $1\hspace{-.5mm} \le m \le\hspace{-.5mm} M$, $0\hspace{-.5mm}\le\beta\hspace{-.5mm}\le1$, and $\varepsilon=10^{-10}$ is a small constant for numerical stability. 
With $\beta=1$, (\ref{eq:PHAT}) becomes the PHAT~\cite{knapp_generalized_1976}, which retains only the phase differences, whereas $\beta=0$ preserves the raw correlations~\cite{shin_tf-corrnet_2025}. We set $\beta=0.5$ by default, which compresses the remaining variations of the correlation magnitudes while preserving their phases. The same weighting is applied to (\ref{eq:corr_multiref}).

\subsection{Multi-channel time-frequency filter estimation}

Conventionally, extending the filtering operation to a multi-tap convolutional form under multi-channel conditions has been shown effective for joint dereverberation and denoising~\cite{wang_tf-gridnet_2023, nakatani_speech_2010, nakatani_dnn-supported_2020, cho_convolutional_2021}. 
Likewise, deep filtering~\cite{mack_deep_2020}, which models adjacent time–frequency bins in a convolution-like manner rather than estimating a single complex mask per bin, has demonstrated strong performance. 

In our formulation, the spatio-spectro-temporal correlations from $\tilde{\mathbf{x}}_{tf}$ with the single reference observation $X_{tf1}$ serve as a feature representation from which the model estimates the corresponding MISO separation filter $\mathbf{w}_{k,tf}\in \mathbb{C}^{1 \times M(2L+1)(2I+1)}$. 
This establishes a structural symmetry between the correlation and the filter: both sharing the same spatio-spectral extent across channels and time-frequency bins. 
Accordingly, the convolutional deep filter $\mathbf{w}_{k,tf}$ for the $k$-th target source is applied to the corresponding spatio-spectro-temporal input $\tilde{\mathbf{x}}_{tf}$ to produce the separated output 
\begin{equation}
    Y_{k,tf} = \mathbf{w}_{k,tf}\tilde{\mathbf{x}}_{tf} \in \mathbb{C}.
\end{equation}

Although the same framework can be extended to a multi-input multi-output (MIMO) setting by estimating outputs for all microphones~\cite{wang_multi-microphone_2021, taherian_leveraging_2024, shin_tf-corrnet_2025}, such a design would substantially increase the correlation dimensionality, as well as the computational and training costs. Therefore, in this study, we focus on the MISO configuration, where the target signal is estimated at the reference channel.

\label{sec:motivation}
The optimal linear multi-channel multi-frame filter depends on the second-order statistics of the observation, i.e., the expectations of the spatio-spectro-temporal correlations in (\ref{eq:corr_MIMO})~\cite{huang_multi-frame_2012, tammen_deep_2021, shin_deep_2026}. The correlations add no information beyond $\tilde{\mathbf{x}}$ itself. They make the pairwise interactions underlying these statistics explicit, providing an inductive bias for the filter output. Deep MVDR filtering estimates the statistics with a network and computes the filter in closed form~\cite{tammen_deep_2021}. SR-CorrNet takes the instantaneous correlations as input and learns a nonlinear mapping to the filter coefficients.

\section{SR-CorrNet}

As illustrated in Fig.~\ref{fig:SR_CorrNet}, SR-CorrNet implements the correlation-to-filter formulation introduced in the previous section\footnote{We omitted the permute and reshape operations in all the figures for simplicity. Instead, to clarify the figure, we indicated the shapes of tensors.}. 
The network takes spatio-spectro-temporal correlations as input, estimates multi-frame filters for each speaker, and reconstructs clean signals through filtering. 
An asymmetric separation-reconstruction architecture, built on a TF dual-path backbone, is employed for feature processing.
This structure allows the model to explicitly encode how speech components correlate across time-frequency and space, while progressively refining separated features in a stage-wise manner.

\begin{figure}
\centering
\subfloat[Correlation module]{\includegraphics[width=0.42\columnwidth]{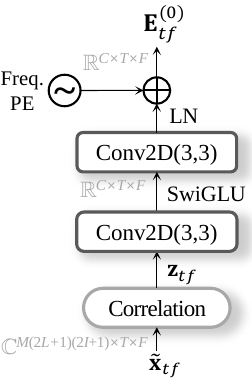}}\hspace{2mm}
\subfloat[Filter module]{\includegraphics[width=0.52\columnwidth]{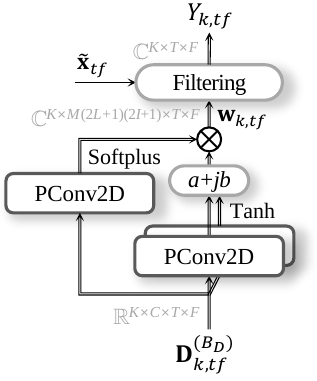}}
\footnotesize
\caption{{Block diagrams of (a) correlation and (b) filter modules. The correlations of (\ref{eq:corr_MISO}) with the PHAT-$\beta$ weighting in (\ref{eq:PHAT}) are flattened to the real-valued vector of $\mathbb{R}^{2M(2L+1)(2I+1)}$.}}
\label{fig:Corr_Filter}
\end{figure}

\subsection{Correlation-to-filter design}

SR-CorrNet adopts a correlation-to-filter paradigm, where the model predicts complex filters directly from spatio-spectro-temporal correlation features rather than mapping noisy spectra to clean ones.
This formulation enables the network to infer separation filters from representations that inherently encode temporal, spectral, and spatial dependencies.
The correlation tensor $\mathbf{z}_{tf}$, computed from multi-frame, multi-band, and multi-channel observations $\tilde{\mathbf{x}}_{tf}$, is used as the primary input feature.

Before entering the encoder, the correlations are embedded by shallow convolutional layers as shown in Fig.~\ref{fig:Corr_Filter}(a).
Two $3\times3$ 2d convolution (Conv2D) layers first project the input correlations to $C$ channels with a gated linear unit with a swish activation (SwiGLU). Then, the input features are normalized by layer normalization (LN) and added by a sinusoidal positional encoding along the frequency axis to help preserve ordering of frequency bins.

At the output, the Filter module estimates a TF complex deep filter for each separated stream. As shown in Fig.~\ref{fig:Corr_Filter}(b), real/imaginary components and additional magnitude masking component are generated by three parallel $1\times 1$ point-wise Conv 2D (PConv2D) layers with an output channel of $M(2L+1)(2I+1)$. Denoting the three outputs by $\mathbf{u}_{k,tf}$, $\mathbf{a}_{k,tf}$, and $\mathbf{b}_{k,tf}\in\mathbb{R}^{1\times M(2L+1)(2I+1)}$, they are combined to form the complex filter as
\begin{equation}
\mathbf{w}_{k,tf} = \mathrm{Softplus}(\mathbf{u}_{k,tf}) \odot \left(\tanh(\mathbf{a}_{k,tf}) + j\tanh(\mathbf{b}_{k,tf})\right),
\label{eq:filter_param}
\end{equation}
where $\odot$ denotes the element-wise product and the activation functions are applied element-wise. These filters are applied to the corresponding $\tilde{\mathbf{x}}_{tf}$ to estimate the separated clean signal as $Y_{k,tf}=\mathbf{w}_{k,tf}\tilde{\mathbf{x}}_{tf}$.

\subsection{Asymmetric encoder-decoder structure}
To estimate the filter from the correlation in the network, the overall architecture consists of a separation encoder, a split module, and a reconstruction decoder, as shown in Fig.~\ref{fig:SR_CorrNet}. 
The $B_E$ stacks of the TF-encoder perform coarse separation by alternately processing features along the frequency and time axes using stacked TF blocks. Each frequency module models frequency sequences within a frame, while each time module captures temporal sequences across frames. This dual-path process allows the encoder to represent both local patterns and long-term dependencies with relatively few parameters.
After the separation encoder stage, the split module projects the latent representation into $K$ speaker-wise feature streams.

The $B_D$ stacks of the TF-decoder reconstruct the separated features $B_D$ times by alternating discriminative learning using weight-sharing TF modules and interaction of separated speaker features using the speaker interaction module~\cite{shin_separate_2024, lee_boosting_2024}. While the weight-shared modules directly learn to capture the discriminative features, because the speech elements can be mistakenly clustered into other speaker streams, the speaker interaction module makes the speaker-wise features attend to each other to recover the distorted components.

\begin{figure}
\centering
\subfloat[Time and frequency modules]{\includegraphics[width=0.58\columnwidth]{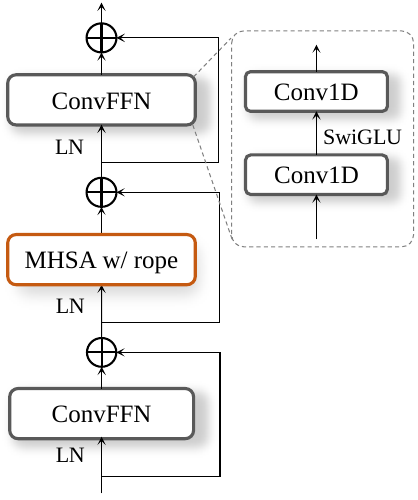}}\hspace{.5mm}
\subfloat[Speaker module]{\includegraphics[width=0.31\columnwidth]{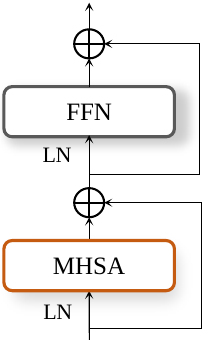}}\hspace{4mm}
\footnotesize
\caption{{Block diagrams of (a) common unit module for time and frequency modules and (b) speaker interaction module.}}
\vspace{-1mm}
\label{fig:Unit_Module}
\end{figure}

\subsection{Unit processing TF-module}

Each TF block used in the TF-encoder and TF-decoder follows the structure shown in Fig.~\ref{fig:Unit_Module}(a).
The time and frequency modules share the common unit module structure~\cite{saijo_tf-locoformer_2024}. They commonly employ multi-head self-attention (MHSA) with rotary positional encoding (RoPE)~\cite{su_roformer_2024}.
As the macaron-style structure, these modules are surrounded by two convolutional feed-forward networks (ConvFFN) that better capture local contexts. 
Each ConvFFN consists of two 1D convolution layers with SwiGLU activation in the hidden layer with a hidden channel dimension of $C_h$. 
All the residual units adopt a pre-norm residual unit~\cite{wang_learning_2019, nguyen_transformers_2019} to facilitate training of deep networks.

In the decoder, the speaker interaction module is based on the vanilla Transformer encoder block~\cite{vaswani_attention_2017, dosovitskiy_image_2021} without positional encoding. 
It performs self-attention across the separated speaker streams for each time-frequency bin component, followed by the basic feed-forward network (FFN), which consists of two linear layers with a hidden dimension of $4C$ and ReLU activation. 
Therefore, the speaker module learns to identify the interfering components of the opposing sequences within the same TF bins.

\subsection{Split module}

The split module is responsible for determining the number of active speakers and dividing the encoded feature into separate streams before reconstruction.
Depending on the task configuration, two different variants are employed: a fixed-speaker version for controlled experiments and a dynamic version based on attractor estimation for real conversational scenarios.

\paragraph{Fixed-speaker split}

Assuming that the number of speakers $K$ is known, the output of TF-Encoder $\mathbf{E}^{(B_E)}_{tf}$ is projected and divided into $K$ streams of $\mathbf{D}^{(0)}_{k,tf}$ using two linear layers with SwiGLU activation in the middle. 
The first linear layer projects the encoder feature with $C$ channels to $KC$ channels. Then, the feature is processed by the second linear layer with the same output channels $KC$ after activation and reshaped to $\mathbb{R}^{K \times T\times F \times C}$, and normalized by LN.
As a simple baseline, this static configuration is mainly used for benchmark datasets with a predefined speaker count.

\paragraph{Attractor-based dynamic split}
In real recordings such as LibriCSS~\cite{chen_continuous_2020},
the number of active speakers varies over time, and long segments
often contain only a single speaker or no speech. Under such conditions, a
static split strategy tends to generate artificial secondary
streams, resulting in \textit{spectral leakage}, where the energy
of one speaker is distributed across multiple outputs.
During continuous speech separation (CSS), processing a fixed number of streams wastes computation on inactive streams, especially in silent segments.
Previous approaches have addressed the leakage by adding a
speaker-counting network~\cite{wang_count_2021} or a
stream-merging stage with localization cues~\cite{taherian_leveraging_2024} to a separator with a fixed number of output streams.
SR-CorrNet integrates speaker counting and speaker conditioning into the separator through a
Transformer-decoder-based attractor module (TDA)~\cite{lee_boosting_2024}
as a dynamic split mechanism for an unknown number of speakers, so that the decoder processes only the streams of the active speakers.
This module allows SR-CorrNet to handle both single-speaker and
multi-speaker regions within a single forward pass by directly
estimating speaker existence probabilities.

\begin{figure}
\footnotesize
\centering
\includegraphics[width=0.92\columnwidth]{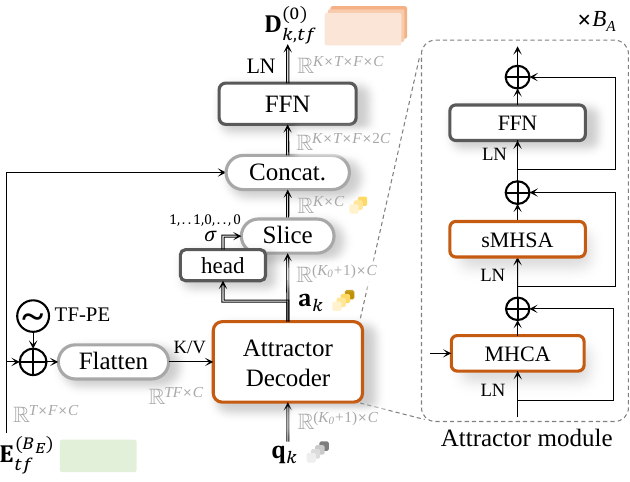}
\caption{
Block diagram of the attractor-based dynamic split module.
The encoder feature is added to 2-d positional embeddings, flattened, and provided as key-value input to the
attractor decoder with learnable query vectors $\mathbf{q}_k$, where the attractor module in the dashed box is stacked $B_A$ times. The attractors $\mathbf{a}_k$ are used to estimate speaker existence
probabilities $p_k$.
}
\label{fig:split_attractor}
\end{figure}

As illustrated in Fig.~\ref{fig:split_attractor}, the dynamic split
module takes the encoder feature $\mathbf{E}_{tf}^{(B_E)}$ and
generates attractors $\mathbf{a}_k \in \mathbb{R}^{C}$, each of
which corresponds to a potential speaker stream. The attractor
decoder is implemented with Transformer decoder blocks
\cite{vaswani_attention_2017, lee_boosting_2024}, using the
flattened encoder feature $\mathbf{E}_{tf}^{(B_E)}$ as the
key-value input and $K_0+1$ learnable query vectors $\mathbf{q}_k \in \mathbb{R}^{C}$ as decoder
queries, where $K_0$ denotes the maximum number of speakers
considered by the model. We stack $B_A$ attractor decoder blocks,
each consisting of multi-head cross-attention (MHCA) followed by masked
MHSA. The causal masking enforces sequential
prediction of speaker existence.
In this paper, we set $B_A = 2$.

Among the resulting attractors, those associated with active
sources are expected to represent valid speaker streams, whereas
the remaining ones correspond to inactive or non-speech slots.
To train this behavior explicitly, each attractor is passed through
an additional linear layer to predict its existence probability
$0 \le p_k \le 1$, and the attractor module is optimized with the
binary cross-entropy (BCE) loss
\begin{equation}
\mathcal{L}_{\text{attr}}
=
\frac{1}{K_0\hspace{-.5mm}+\hspace{-.5mm}1}
\hspace{-.5mm}\left(
\sum_{k=1}^{K}\text{BCE}(p_k, 1)
+
\hspace{-2mm}
\sum_{k=K+1}^{K_0+1}
\hspace{-2mm}
\text{BCE}(p_k, 0)
\hspace{-.5mm}
\right)\hspace{-.5mm},
\end{equation}
where the first $K$ attractors are assigned to active speakers and
the remaining attractors are trained to indicate non-existence.
Since $\mathcal{L}_{\text{attr}}$ depends only on the number of active speakers, it is invariant to the permutation of the output streams, which is determined by PIT in the separation loss.

For feature splitting, the first $K$ attractors are selected and
broadcast along the time and frequency axes, then concatenated
with the separation TF-encoder feature $\mathbf{E}_{tf}^{(B_E)}$ to form a speaker-conditioned feature
tensor in $\mathbb{R}^{K \times T \times F \times 2C}$. This
tensor is then passed through two linear layers with hidden SwiGLU activation
to produce $K$ active feature streams
$\mathbf{D}_{k,tf}^{(0)}$.
At inference, we count the attractors whose existence probabilities exceed 0.5, cap the estimated speaker count $\hat{K}$ at $K_0$, and use the first $\hat{K}$ attractors for splitting. When $\hat{K}=0$, we save computation by skipping the decoder and outputting zero signals. Even when more than $K_0$ speakers are active, the separator outputs at most $K_0$ streams.

\subsection{Boosting SepRe method by early supervision}

Because the feature sequences are already separated through the decoder stages, 
each reconstruction decoder block $B_D$ can be independently trained with a stage-specific separation objective as a form of early supervision.
From the separated feature from the split module $\mathbf{D}^{(0)}_{k,tf}$ and the output features of $B_D$ decoder blocks $\mathbf{D}^{(b)}_{k,tf}, b=1,...,B_D$, the source signals can be estimated as 
\begin{equation}
\hat{Y}^{(b)}_{k,tf} = \hat{\mathbf{w}}^{(b)}_{k,tf}\tilde{\mathbf{x}}_{tf},   
\end{equation}
where $\hat{\mathbf{w}}^{(b)}_{k,tf}$ represents the auxiliary ($0\le b\le B_D-1$) and final filter ($b=B_D$) derived from the intermediate or final decoder feature $\mathbf{D}^{(b)}_{k,tf}$.

The final output is obtained as $Y_{k,tf}=\hat{Y}^{(B_D)}_{k,tf}$, 
and the corresponding main objective function for speech separation is denoted by $\mathcal{L}_{B_D}$.
For the auxiliary outputs $\hat{Y}^{(b)}_{k,tf}, 0\le b \le B_D-1$, we define an auxiliary loss $\mathcal{L}_b$ for each intermediate stage.
Accordingly, the total loss for boosting the SepRe method is formulated as
\vspace{-1mm}
\begin{equation}
\mathcal{L}_\text{fix} = (1-\alpha)\mathcal{L}_{B_D} + \alpha \sum^{B_D-1}_{b=0}\frac{\mathcal{L}_b}{B_D},
\label{multi_loss_fix}
\end{equation}
where $\alpha$ is a balancing coefficient.
If we use the dynamic split module, the final loss can be set to 
\vspace{-1mm}
\begin{equation}
\mathcal{L}_\text{dyn} = (1-\alpha)\mathcal{L}_{B_D} + \alpha \sum^{B_D-1}_{b=0}\frac{\mathcal{L}_b}{B_D} + \mathcal{L}_\text{attr}
\end{equation}
by considering the attractor existence probability loss.

Since the auxiliary outputs are used only during training, 
the objective function for $\mathcal{L}_b, 0\le b\le B_D-1$ can adopt a more relaxed form for stable optimization, for instance, a distance measure between magnitude spectra in the STFT domain without enforcing phase consistency. 

\section{Experimental Setup}
To evaluate the proposed SR-CorrNet\footnote{The code for the speech separation experiments on the standard benchmarks is available at \url{https://github.com/dmlguq456/SR_CorrNet_SS}.}, we experimented on simulated and real-recorded mixtures of various numbers of speakers with single- and multi-channel observations.

\subsection{Monaural speech separation for anechoic clean mixture}
\label{sec:setup_wsj}
\paragraph{Dataset and evaluation metrics} We used WSJ0-2/3/4/5mix~\cite{hershey_deep_2016, isik_single-channel_2016, nachmani_voice_2020} as the most basic and popular dataset to benchmark single-channel anechoic speech separation, which have 20,000, 5,000 and 3,000 mixtures for each number of speakers in its training, validation, and test sets where the speakers are also partitioned. The source signals were fully overlapped in an utterance, and their power levels were uniformly sampled from the range $[-5,5]$dB. We used the `min' configuration and a sampling rate of 8kHz. As evaluation metrics, we used scale-invariant signal-to-noise ratio improvement (SI-SNRi)~\cite{luo_conv-tasnet_2019} and signal-to-distortion ratio improvement (SDRi)~\cite{roux_sdr_2019}.
\paragraph{Model Configuration and Training}
As the task does not contain reverberations, the TF context window for correlations and filter estimation was set to $(L, I)=(1,1)$, which is equivalent to $(2L+1, 2I+1)=(3,3)$. STFT was computed using a Hanning window with a length of 128 samples and a hop of 64 samples. We tested three model configurations:
\begin{itemize}
\small
    \item SR-CorrNet-S: $C=96, C_h=256, B_E=2, B_D=4$
    \item SR-CorrNet-B: $C=128, C_h=384, B_E=2, B_D=4$
    \item SR-CorrNet-L: $C=192, C_h=512, B_E=3, B_D=5$
\end{itemize}
The number of heads in the TF and speaker modules and the kernel size in the TF module were set to 4 and 3, respectively. For evaluation of WSJ0-2mix alone, we used fixed split module ($K=2$) while dynamic split was employed on SR-CorrNet-B for the extended (WSJ0-2/3/4/5mix) experiment. 

The models were trained with permutation invariance training (PIT)~\cite{kolbaek_multitalker_2017}. For training, 4-s utterance segments were used with a batch size of 1. For the objective function, we used SI-SNR~\cite{luo_conv-tasnet_2019} as
\begin{equation}
\mathcal{L}_{B_D} = -\sum_{k=1}^K 20\log_{10}{\frac{\|\gamma_k{s}_k\|_2}{\|\gamma_k{s}_k-{y}_k\|_2}},
\end{equation}
where $s_k$ and $y_k$  are the $k$-th source signal and the corresponding output based on PIT, and $\gamma_k={{y}}^T_k{s}_k/\|{s}_k\|_2^2$. Also, a clipping was applied at 30 dB to limit the influence of the best training prediction~\cite{zeghidour_wavesplit_2021}. For the auxiliary function of $\mathcal{L}_{b}, 0\le b\le B_D-1$, we used the magnitude values of $s_k$ and $y_k$ as
\begin{equation}
\mathcal{L}_{b} = -\sum_{k=1}^K 20\log_{10}{\frac{\|{|\mathcal{F}({\gamma_ks}}_k)|\|_2}{\||\mathcal{F}(\gamma_k{s}_k)|-|\mathcal{F}(y_k)|\|_2}},
\end{equation}
where $\mathcal{F}(\cdot)$ is the STFT operator and $\alpha$ in (\ref{multi_loss_fix}) was set to 0.5 and decayed by a factor of 0.95 at every epoch after 30 epochs. The models were trained up to 100 epochs by the AdamW optimizer~\cite{loshchilov_decoupled_2019} with a weight decay of 0.01 and a warm-up training scheduler for the first 5000 steps up to an initial learning rate of $1.0e^{-3}$. The learning rate decayed by a factor of 0.95 at every epoch after 50. Gradient clipping was applied with a maximum L2-norm of 5. 
For the SR-CorrNet-L, we applied dynamic mixing (DM)~\cite{zeghidour_wavesplit_2021, subakan_attention_2021} and the initial learning rate was set to $2.0e^{-4}$ and decayed after 50 epochs.

\subsection{Speech separation for simulated noisy-reverberant mixtures}
\label{sec:setup_whamr}
\paragraph{Dataset and evaluation metrics}
To complement unrealistic anechoic clean mixing of WSJ0-mix, we used WHAMR! dataset~\cite{maciejewski_whamr_2020}, which introduces stereo reverberation for each source in addition to background noise~\cite{wichern_wham_2019} based on the same recipe of WSJ0-2mix.
Its dataset partitions are the same as those of WSJ0-2mix.
In each mixture, the reverberation time ($T_{60}$) and speaker-to-array distance were randomly sampled from the range $[0.2, 1.0]$ s and from $[0.66, 2.0]$ m, respectively.
The background noise was added with SNR between the louder speaker and noise from $[-6,3]$ dB. As in WSJ0-mix, we used its `min' and 8kHz version, and SI-SNRi and SDRi as evaluation metrics. For both the single- and two-channel settings, the anechoic source signals at the first microphone provided by the dataset were used as the targets for training and evaluation, as in~\cite{saijo_tf-locoformer_2024, wang_tf-gridnet_2023, quan_spatialnet_2024}.
\paragraph{Model Configuration and Training}
Considering the reverberation, the context window of TF bins for $\tilde{\mathbf{x}}_{tf}$ was set to $(L, I)=(3,1)$, which is equivalent to $(2L+1, 2I+1)=(7,3)$.
The channel dimension is set to $C=128$, and $B_E$ and $B_D$ were set to 2 and 4, respectively, as in SR-CorrNet-B from Subsection~\ref{sec:setup_wsj}. As in previous studies~\cite{wang_tf-gridnet_2023, saijo_tf-locoformer_2024}, we used a longer kernel size of 7 in ConvFFN with reduced $C_h=256$ and a longer STFT window of 256 samples with a hop of 64 samples, with all elements of $\tilde{\mathbf{x}}_{tf}$ divided by the reference-channel RMS at the $f$-th frequency. A fixed split module with $K=2$ was used, and the training procedure was the same as in Subsection~\ref{sec:setup_wsj}.
Two training runs of SR-CorrNet-B and SR-CorrNet-L + DM for Tables~\ref{tab:WSJ2mix}--\ref{tab:WHAMR!_2ch} differed by less than 0.1~dB; thus, standard deviations are omitted.

\begin{figure}
\footnotesize
\centering
\includegraphics[width=0.94\columnwidth]{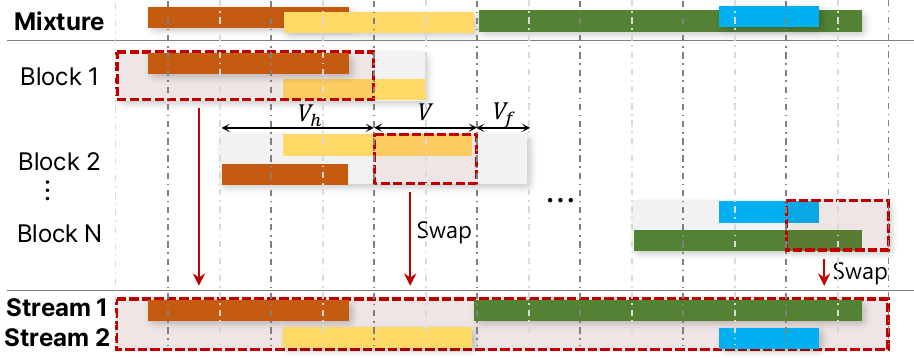}
\vspace{-1mm}
\caption{{Illustration of CSS scheme. In our experiment, we set the chunk-size of $V_{h} = 1.2\ s, V = 0.8\ s$, and $V_f = 0.4\ s$, respectively.}}
\label{fig:CSS}
\vspace{-4mm}
\end{figure}

\subsection{Continuous speech separation for real-recorded mixtures}
\label{sec:setup_css}
\paragraph{Dataset and evaluation metrics}
As real-recorded noisy-reverberant multi-channel mixtures, we used LibriCSS~\cite{chen_continuous_2020}, a 10-hour 7-channel dataset recorded in meeting-like scenario based on the Librispeech test set, and the AMI meeting corpus~\cite{carletta_ami_2005}, which consists of real meetings with spontaneous conversations recorded with an eight-microphone circular array. For LibriCSS, the evaluation schemes were twofold: utterance-wise evaluation and continuous input evaluation. Because our focus was mainly on speech separation, we utilized an ASR model used in the original LibriCSS paper~\cite{chen_continuous_2020} for evaluation. For AMI, we used the official development and evaluation sets, consisting of 18 and 16 meetings, respectively. The output streams were transcribed without adaptation by off-the-shelf recognizers, Parakeet-TDT~\cite{xu_efficient_2023}, Whisper large-v3~\cite{radford_robust_2023}, and Canary-1B~v2~\cite{sekoyan_canary-1b-v2_2025}, on the same segments of at most 25~s covering each meeting, and the same text normalization~\cite{radford_robust_2023} was applied to the references and the hypotheses. Since these recognizers tend to generate words even for nearly silent inputs, an output stream was transcribed in a segment only when it contained voice activity longer than 0.5~s. We report the optimal reference combination WER (ORC-WER)~\cite{sklyar_multi-turn_2022, von_neumann_meeteval_2023}, in which each reference utterance is assigned to one of the output streams so as to minimize the total number of errors. Following the previous studies~\cite{chen_continuous_2020, chen_continuous_2021, wang_multi-microphone_2021, shin_tf-corrnet_2025}, we used the same chunk-wise online processing and stitching with the same configurations as illustrated in Fig.~\ref{fig:CSS} for both datasets. We process each chunk with its history and future contexts. We align the separated streams with those of the previous chunk using the permutation that minimizes their difference in the overlapping region, then emit only the center segment.

\paragraph{Model Configuration and Training}
Instead of using longer frames, the TF context window for correlation was set to $(L, I)=(2,1)$, which is equivalent to $(2L+1, 2I+1)=(5,3)$, considering the sufficient number of microphones ($M=7$ for LibriCSS and $M=8$ for AMI), and to $(L,I)=(3,1)$ for the single-channel models. The other configuration was the same as SR-CorrNet-S in Subsection~\ref{sec:setup_wsj} except that the $B_E$ and $B_D$ were commonly set to 3 to further reduce the computations. For the split module, we tested the fixed split module ($K=2$) and the dynamic split module with $K_0=3$ ($K_0=2$ for the single-channel model), which outputs $K_0$ streams with zero signals for inactive speakers (denoted by SR-CorrNet-\textit{fix}\textsubscript{2} and SR-CorrNet-\textit{var}\textsubscript{0-$K_0$}, respectively). 

To train the networks for evaluation of LibriCSS and AMI, we simulated reverberated 7-channel and 8-channel mixtures, respectively, with the microphone array of each dataset using the gpuRIR~\cite{diaz-guerra_gpurir_2021} with $T_{60} \in [0.2$s$,~0.7$s$]$ based on the Librispeech dataset~\cite{panayotov_librispeech_2015}, and no real recordings of either dataset were used for training. The sampling rate was 16 kHz. In this case, we considered various mixing configurations, as in~\cite{yoshioka_recognizing_2018} with an average overlap ratio of about 50\%. We added two types of noises with a signal-to-noise ratio (SNR) ranging from 5 to 20 dB, respectively: spatially diffuse random colored white noise and reverberant point-source noise by convolving a simulated RIR and noise samples from DNS challenge dataset~\cite{reddy_interspeech_2020}. The learning rate was set to $2.0e^{-4}$ and reduced by a factor of 0.9 for every 20k steps after 400k steps up to 1000k steps with a batch size of 2. 
The utterance length for training was set to the corresponding chunk size $V_h+V+V_f = (1.2 + 0.8 +0.4)$~s.

For training loss, we used L1 loss on magnitude and complex components as
\begin{equation}
\mathcal{L}_{B_D} \hspace{-.5mm}= \hspace{-.5mm}\sum_{k=1}^K\hspace{-1mm}\left(\mathcal{L}_{\text{M}}(y_k,s_k)+\frac12\mathcal{L}_{\text{R}}(y_k,s_k)+\frac12\mathcal{L}_{\text{I}}(y_k,s_k)\right),
\end{equation}
where
\begin{eqnarray}
\hspace{-5mm}    \mathcal{L}_{\text{M}}(y_k,s_k) \hspace{-2mm}&=&\hspace{-2mm} \left\||\mathcal{F}(y_k)|-|\mathcal{F}(s_k)|  \right\|_1,\\
\hspace{-5mm}    \mathcal{L}_{\text{R}}(y_k,s_k)\hspace{-2mm} &=&\hspace{-2mm} \left\|\mathrm{Real}\left(\mathcal{F}(y_k)\right)-\mathrm{Real}\left(\mathcal{F}(s_k)\right)  \right\|_1,\\
\hspace{-5mm}    \mathcal{L}_{\text{I}}(y_k,s_k)\hspace{-2mm} &=&\hspace{-2mm} \left\|\mathrm{Imag}\left(\mathcal{F}(y_k)\right)-\mathrm{Imag}\left(\mathcal{F}(s_k)\right)  \right\|_1.
\end{eqnarray}
The auxiliary losses for the SepRe method were used as
\begin{equation}
\mathcal{L}_{b} \hspace{-.5mm}= \hspace{-.5mm}\sum_{k=1}^K\hspace{-0mm}\mathcal{L}_{\text{M}}(y_k,s_k).
\end{equation}
As in Section~\ref{sec:setup_wsj}, the model was trained based on PIT~\cite{kolbaek_multitalker_2017}.
Also, to stabilize the training and minimize the unnecessary distortions of speech signals, we set the target source $s_k$ as convolution of the truncated RIR filter and the corresponding source signal. In the RIR filter, the $n_\text{peak}+n_\text{offset}$ samples were used where $n_\text{peak}$ is peak index from the direct-path component and $n_\text{offset}$ is a tunable offset. We empirically set $n_\text{offset}=256$ for the multi-channel case of LibriCSS and $n_\text{offset}=512$ for the single-channel case and for AMI.

\paragraph{Direction-based stitching}
The similarity-based stitching compares the separated signals in the overlapping region and is thus sensitive to the changes of the active speakers within a chunk. Prior work used speaker directions to group separated streams across chunks~\cite{wang_localization_2022} and merge streams from a single speaker~\cite{taherian_leveraging_2024}. These methods estimate directions with GCC-PHAT~\cite{knapp_generalized_1976} from the signals that the separation networks estimate for all microphones, requiring a MIMO network or repeated MISO inferences. We keep the MISO configuration and attach an auxiliary head to the decoder output of each stream, which outputs the direction of the speaker as a two-dimensional unit vector at every frame and is trained with the cosine distance to the azimuth of each simulated source in the frames where the source is active. The direction of each output is obtained as the energy-weighted circular mean of its frame outputs over the center segment. For the stitching, each stream keeps the direction of the speaker it carried last, and the permutation that matches the largest number of outputs of a new chunk to the stream directions within a matching angle is selected. In addition, when the directions of two outputs of a chunk are within the matching angle, they are regarded as the same speaker and the output with the lower energy is added to the other one as in~\cite{taherian_leveraging_2024}, which complements the dynamic split when a single speaker is still separated into two streams.

\section{Evaluation Results}

\subsection{Monaural speech separation for anechoic clean mixtures}

\begin{table}
\footnotesize
\caption{Evaluation on WSJ0-2mix dataset. Results of the compared systems are taken from the literature. $\dagger$ denotes the use of additional speaker information.}
\renewcommand{\tabcolsep}{6.0pt}
\def\arraystretch{1.05}
\centering
\begin{tabular}{lcccc}
\toprule
\multirow{2}{*}{\textbf{System}} & \textbf{Params.} & \textbf{MACs} & \textbf{SI-SNRi} & \textbf{SDRi} \\
 & (M) & (G/s) & (dB) & (dB) \\
\midrule
Conv-TasNet~\cite{luo_conv-tasnet_2019}& 5.1 & 10.5 &15.3 & 15.6\\
SuDoRM-RF~\cite{tzinis_sudo_2020}& 6.4 & 10.1  &18.9&-\\
DPRNN~\cite{luo_dual-path_2020}& 2.6 & 88.5 &18.8&19.0\\
DPTNet~\cite{chen_dual-path_2020}&2.7&102.5&20.2&20.3\\
Sepformer~\cite{subakan_attention_2021}&26.0&86.9&20.4&20.5\\
A-FRCNN~\cite{hu_speech_2021}&6.1&125.0&18.3&18.6\\
SFSRNet~\cite{rixen_sfsrnet_2022} & 59.0 & 124.2 & 22.0 & 22.1\\
ISCIT~\cite{mu_multi-dimensional_2023}&58.4& 252.2 & 22.4 & 22.5 \\
QDPN~\cite{rixen_qdpn_2022} & 200.0 & - & 22.1 & - \\
TF-GridNet~\cite{wang_tf-gridnet_2023}& 14.5 & 460.8 & 23.5 & 23.6 \\
SepReformer-T~\cite{shin_separate_2024} &3.5&10.4& 22.4 & 22.6  \\
SepReformer-B~\cite{shin_separate_2024} &14.2&39.8&{23.8}& {23.9} \\
TF-Locoformer-M~\cite{saijo_tf-locoformer_2024} & 15.0 & 204.3 & 23.6 & 23.8 \\
TF-Locoformer-L~\cite{saijo_tf-locoformer_2024} & 22.5 & 306.2 & \bf 24.2 & \bf 24.3 \\
\rowcolor{Gray}
SR-CorrNet-S & 7.0 & 88.9 & 23.6 & 23.8\\
\rowcolor{Gray}
SR-CorrNet-B & 13.6 & 177.3 & 24.1 & \bf 24.3 \\
\midrule
Sepformer + DM~\cite{subakan_attention_2021}&26.0&86.9&22.3&22.5\\
SFSRNet + DM~\cite{rixen_sfsrnet_2022}&59.0&466.2&24.0&24.1\\
ISCIT$^\dagger$ + DM~\cite{mu_multi-dimensional_2023}&58.4& 252.2 & 24.3 & 24.4 \\
QDPN + DM~\cite{rixen_qdpn_2022} & 200.0 & - & 23.6 & - \\
{SepReformer-L + DM~\cite{shin_separate_2024}\hspace{-2mm}} & 59.4 & 155.5 & 25.1 & 25.2\\
TF-Locoformer-L + DM~\cite{saijo_tf-locoformer_2024} & 22.5 & 306.2 & 25.1 & 25.2 \\
\rowcolor{Gray}
SR-CorrNet-L + DM & 36.0 & 467.6 & \bf 25.5 & \bf 25.7 \\
\bottomrule
\end{tabular}
\vspace{-2mm}

\label{tab:WSJ2mix}
\end{table} 

\paragraph{Comparison with existing models on WSJ0-2mix}  
Table~\ref{tab:WSJ2mix} presents the evaluation results on the WSJ0-2mix dataset.  
The proposed SR-CorrNet shows competitive performance compared with recent mapping-based architectures such as SepReformer and TF-Locoformer, confirming the effectiveness of correlation-based deep filtering on clean, anechoic mixtures.  
Without DM, SR-CorrNet-S achieved the same SI-SNRi as TF-Locoformer-M, a similar TF-domain backbone, with less than half of its MACs, and SR-CorrNet-B further improved the performance. Time-domain models such as SepReformer require fewer MACs, whereas SR-CorrNet prioritizes separation accuracy within the TF domain, which becomes advantageous under noisy-reverberant conditions as shown in Section~\ref{sec:whamr_result}.
When trained with DM, SR-CorrNet-L further improved the separation accuracy and achieved the best overall performance among all compared systems\footnote{The result of SepReformer-L + DM is taken from the latest arXiv version (arXiv:2406.05983v4) of~\cite{shin_separate_2024}, which reports the corrected value of its Table~5(b).}, at the cost of higher MACs as an accuracy-oriented configuration, while the S and B models provide favorable quality--cost trade-offs among the TF-domain models.  
Although the correlation–input and filtering–output design was originally motivated by multi-channel and reverberant conditions, these results verified that it remained highly effective even in anechoic monaural environments, suggesting its potential as a general and unified separation framework.

\begin{table}
\footnotesize
\caption{{Evaluation of SI-SNRi (dB) results on WSJ0-$K$mix dataset. ``Prior $K$" denotes whether the number of speakers in the mixture is given in advance, and the speaker counting accuracy (\%) is given in parentheses. Results of the compared systems are taken from the literature.}}
\renewcommand{\tabcolsep}{2.0pt}
\def\arraystretch{1.04}
\centering
\footnotesize
\scalebox{0.95}{\begin{tabular}{lccllll}
\toprule
\multirow{2}[2]{*}{\textbf{System}} & \multirow{1}[2]{*}{\textbf{Split}} & \multirow{1}[2]{*}{\textbf{Prior}} & \multicolumn{4}{c}{$K$} \\
\cmidrule{4-7}
& \multirow{1}[0]{*}{\textbf{Type}} & \textbf{$K$} & \multicolumn{1}{c}{\textbf{2}} & \multicolumn{1}{c}{\textbf{3}} & \multicolumn{1}{c}{\textbf{4}} & \multicolumn{1}{c}{\textbf{5}} \\
\midrule
Sepformer~\cite{subakan_attention_2021}&F&Yes&20.4&17.6&\multicolumn{1}{c}{-}&\multicolumn{1}{c}{-}\\
WaveSplit~\cite{zeghidour_wavesplit_2021}&F&Yes&21.0&17.3&\multicolumn{1}{c}{-}&\multicolumn{1}{c}{-}\\
Mossformer(S)~\cite{zhao_mossformer_2023}&F&Yes&20.9&17.8&\multicolumn{1}{c}{-}&\multicolumn{1}{c}{-}\\
Mossformer(L)+DM~\cite{zhao_mossformer_2023}&F&Yes&22.8&21.2&\multicolumn{1}{c}{-}&\multicolumn{1}{c}{-}\\
SepEDA\textsubscript{2/3}~\cite{chetupalli_speech_2022}&D&Yes&21.5&19.9&\multicolumn{1}{c}{-}&\multicolumn{1}{c}{-}\\
SepTDA\textsubscript{2/3}~\cite{lee_boosting_2024}&D&Yes&{24.0}&\bf 23.7&\multicolumn{1}{c}{-}&\multicolumn{1}{c}{-}\\
\rowcolor{Gray}
SR-CorrNet-B\textsubscript{2/3}&D&Yes&\bf 24.2&23.2&\multicolumn{1}{c}{-}&\multicolumn{1}{c}{-}\\
\midrule
SepTDA\textsubscript{2/3}~\cite{lee_boosting_2024}&D&No&24.0\textsubscript{(100.0)}&22.8\textsubscript{(97.9)}&\multicolumn{1}{c}{-}&\multicolumn{1}{c}{-}\\
\rowcolor{Gray}
SR-CorrNet-B\textsubscript{2/3}&D&No&\textbf{24.2}\textsubscript{(100.0)}&\textbf{23.2}\textsubscript{(99.8)}&\multicolumn{1}{c}{-}&\multicolumn{1}{c}{-}\\
\midrule
Recursive SS~\cite{takahashi_recursive_2019}&D&Yes&14.8&12.6&10.2&\multicolumn{1}{c}{-}\\
Gated DPRNN~\cite{nachmani_voice_2020}&F&Yes&20.1&16.9&12.9&10.6\\
SepEDA\textsubscript{[2-5]}~\cite{chetupalli_speech_2022}&D&Yes
&21.1&18.6&14.7&12.1\\
SepTDA\textsubscript{[2-5]}~\cite{lee_boosting_2024}&D&Yes
&23.6&23.5&22.0&\bf 21.0\\
\rowcolor{Gray}
SR-CorrNet-B\textsubscript{[2-5]}&D&Yes&\bf 24.8&\bf 24.5&\bf 22.1&20.4\\
\midrule
Gated DPRNN~\cite{nachmani_voice_2020}&F&No&18.6&14.6&11.5&10.4\\
SepEDA\textsubscript{[2-5]}~\cite{chetupalli_speech_2022}&D&No
&21.1\textsubscript{(99.8)}&18.4\textsubscript{(97.0)}&14.4\textsubscript{(90.2)}&11.6\textsubscript{(\textbf{96.9})}\\
SepTDA\textsubscript{[2-5]}~\cite{lee_boosting_2024}&D&No
&23.6\textsubscript{(99.9)}&22.1\textsubscript{(95.9)}&19.5\textsubscript{(90.1)}&16.9\textsubscript{(83.2)}\\
\rowcolor{Gray}
SR-CorrNet-B\textsubscript{[2-5]}&D&No
& \textbf{24.8}\textsubscript{(\textbf{100.0})}&\textbf{24.4}\textsubscript{(\textbf{99.7})}&\textbf{21.9}\textsubscript{(\textbf{97.7})}&\textbf{19.9}\textsubscript{(\textbf{96.9})}\\

\bottomrule
\end{tabular}}
\vspace{-2mm}

\label{tab:Kmix}
\end{table} 

\paragraph{Evaluation on WSJ0-$K$mix with dynamic split}  
Table~\ref{tab:Kmix} reports the results on WSJ0-\{2/3/4/5\}mix datasets, where the proposed SR-CorrNet was evaluated under dynamic split conditions and compared with both fixed-$(F)$ and dynamic-$(D)$ separation systems.  
For two- and three-speaker mixtures, SR-CorrNet-B\textsubscript{2/3} achieved the best performance in most conditions and outperformed all compared methods when the number of speakers was unknown.  
Notably, the performance gap between the known and unknown speaker-count conditions remained marginal, indicating that SR-CorrNet reliably inferred the number of active sources and preserved separation accuracy even without prior information.  
When extended to mixtures containing up to five speakers, SR-CorrNet-B\textsubscript{[2–5]} maintained stable performance as $K$ increased without the prior count, while SepTDA\textsubscript{[2–5]} exhibited a relatively larger variation.  
For five-speaker mixtures, SepTDA\textsubscript{[2–5]} achieved higher SI-SNRi with the oracle speaker count. Without the prior count, SR-CorrNet achieved substantially higher SI-SNRi and counted speakers more accurately.  
This suggests that SR-CorrNet generalizes well to mixtures with varying numbers of sources.

\subsection{Speech separation under noisy-reverberant conditions}
\label{sec:whamr_result}

\begin{table}
\footnotesize
\vspace{-2mm}
\caption{{Evaluation on WHAMR! (1ch)}. Results of the compared systems are taken from the literature. $\dagger$ denotes the use of additional speaker information.}
\renewcommand{\tabcolsep}{2.4pt}
\def\arraystretch{1.05}
\centering
\footnotesize
\begin{tabular}{lcccc}
\toprule
\multirow{2}{*}{\textbf{System}} & \textbf{Params.} & \textbf{MACs} & \textbf{SI-SNRi} & \textbf{SDRi} \\
 & (M) & (G/s) & (dB) & (dB) \\
\midrule
\multicolumn{5}{c}{\textit{Time-domain}}\\
WaveSplit + DM$^\dagger$~\cite{zeghidour_wavesplit_2021} & 29.0 & - & 13.2 & 12.2\\
SuDoRM-RF + DM~\cite{tzinis_sudo_2020} & - & - & 13.5 & -\\
Sepformer + DM~\cite{subakan_exploring_2023} & 25.7 & 86.9 & 14.0 & 13.0\\
QDPN + DM~\cite{rixen_qdpn_2022} & 200.0 & - & 14.4 & - \\
MossFormer2 + DM~\cite{zhao_mossformer2_2023} & 55.7 & - & 17.0 & - \\
SepReformer-L + DM~\cite{shin_separate_2024} & 59.4 & 155.5 & 17.2 & 16.0\\
\midrule
\multicolumn{5}{c}{\textit{TF-domain}}\\
U-Net + TCN (SISO$_1$)~\cite{wang_multi-microphone_2021}, \cite{wang_tf-gridnet_2023} & 6.9 & - & 10.3 & 9.7 \\
TF-GridNet (DNN$_1$)~\cite{wang_tf-gridnet_2023} & 5.5 & - & 16.7 & 15.2 \\
TF-GridNet (DNN$_1$+WF+DNN$_2$)~\cite{wang_tf-gridnet_2023} & - & - & 17.3 & 15.8 \\
TF-CrossNet (DNN$_1$)~\cite{kalkhorani_tf-crossnet_2024} & 6.6 & 96.1 & 17.9 & 16.4 \\
TF-CrossNet (DNN$_1$+WF+DNN$_2$)~\cite{kalkhorani_tf-crossnet_2024} & - & - & 18.1 & 16.6 \\
TF-Locoformer-M~\cite{saijo_tf-locoformer_2024} & 15.0 & - & 18.5 & 16.9 \\
FLA-TFLocoformer-M~\cite{wang_flasepformer_2025} & - & - & 18.7 & 17.0 \\
\rowcolor{Gray}
SR-CorrNet-B & 14.8 & 382.4 & \bf 19.7 & \bf 18.1 \\
\bottomrule
\end{tabular}
\label{tab:WHAMR!}
\end{table} 

\begin{table}
\footnotesize
\vspace{-2mm}
\caption{{Evaluation on WHAMR! (2ch)}. Results of the compared systems are taken from the literature.}
\renewcommand{\tabcolsep}{2.8pt}
\def\arraystretch{1.05}
\centering
\footnotesize
\begin{tabular}{lcccc}
\toprule
\multirow{2}{*}{\textbf{System}} & \textbf{Params.} & \textbf{MACs} & \textbf{SI-SNRi} & \textbf{SDRi} \\
 & (M) & (G/s) & (dB) & (dB) \\
\midrule
\multicolumn{5}{c}{\textit{Time-domain}}\\
MC-TasNet~\cite{zhang_time-domain_2021} & - & - & 12.1 & -\\
MC-TasNet w/ spk. extraction~\cite{zhang_time-domain_2021} & - & - & 13.4 & -\\
\midrule
\multicolumn{5}{c}{\textit{TF-domain}}\\
TF-GridNet (DNN$_1$)~\cite{wang_tf-gridnet_2023} & - & - & 18.6 & 17.0 \\
TF-GridNet (DNN$_1$+WF+DNN$_2$)~\cite{wang_tf-gridnet_2023} & - & - & 19.8 & 18.3 \\
SpatialNet~\cite{quan_spatialnet_2024} & 6.5 & 119.0 & 20.2 & 18.5\\
\rowcolor{Gray}
SR-CorrNet-B & 15.0 & 385.7 & \bf 21.8 & \bf 20.1 \\
\bottomrule
\end{tabular}\vspace{-2mm}
\label{tab:WHAMR!_2ch}
\end{table} 

\paragraph{Evaluation under monaural condition}  
Table~\ref{tab:WHAMR!} presents the separation results on the WHAMR! dataset under the monaural condition, which involves simultaneous noise and reverberation, providing a realistic and challenging scenario for joint denoising, dereverberation, and separation.  
Time-domain frameworks show a noticeable performance drop in this setting.  
For example, although SepReformer-L+DM performs competitively on WSJ0-2mix, its score decreases to 17.2\,dB in SI-SNRi, suggesting a limitation of purely time-domain approaches that must implicitly handle both additive and convolutive distortions within a short receptive field.  
In contrast, recent TF-domain models exhibit stronger robustness by explicitly modeling spectral and temporal contexts.  
Among them, SR-CorrNet achieved 19.7\,dB SI-SNRi, surpassing TF-Locoformer~\cite{saijo_tf-locoformer_2024} and its recent variant FLA-TFLocoformer~\cite{wang_flasepformer_2025}.  
This result demonstrates the effectiveness of the proposed SR-CorrNet architecture, where integrating the SepRe framework and correlation-based filtering within a TF-domain model leads to more coherent and robust separation under noisy-reverberant conditions.

\paragraph{Evaluation under stereo condition}  
Table~\ref{tab:WHAMR!_2ch} shows that the stereo SR-CorrNet further improved the performance to 21.8\,dB SI-SNRi, outperforming both TF-GridNet~\cite{wang_tf-gridnet_2023} and SpatialNet~\cite{quan_spatialnet_2024}.  
This result indicates that applying the SR-CorrNet framework to stereo inputs effectively leverages inter-channel correlations in the TF domain without the need for explicit beamforming.  

Overall, the results reveal a clear contrast between WSJ0-2mix and WHAMR! conditions.  
While time-domain approaches perform adequately in clean mixtures, they show substantial degradation in realistic acoustic environments.  
In contrast, SR-CorrNet maintains high separation quality across both monaural and stereo conditions, verifying that the combination of TF-domain modeling, correlation-based filtering, and the SepRe framework is robust to noisy-reverberant mixtures.

\begin{table}
\footnotesize
\vspace{-2mm}
\caption{Ablation of the SepRe framework on WHAMR! (SI-SNRi, dB). Params. and MACs are for the single-channel models.}
\renewcommand{\tabcolsep}{3.5pt}
\def\arraystretch{1.05}
\centering
\footnotesize
\begin{tabular}{lcccc}
\toprule
\textbf{Configuration} & \textbf{Params.} (M) & \textbf{MACs} (G/s) & \textbf{1ch} & \textbf{2ch} \\
\midrule
\rowcolor{Gray}
SepRe ($B_E$, $B_D$) = (2,4) & 14.8 & 382.4 & \bf 19.7 & \bf 21.8 \\
\ w/o early supervision ($\alpha=0$) & 14.8 & 382.4 & 19.2 & 21.1 \\
\ w/o speaker interaction & 13.8 & 349.3 & 19.1 & 21.4 \\
SepRe ($B_E$, $B_D$) = (3,3) & 14.6 & 341.1 & 19.4 & 21.5 \\
Enc.-only ($B_E$, $B_D$) = (6,0) & 13.8 & 217.1 & 18.3 & 20.2 \\
\bottomrule
\end{tabular}
\vspace{-2mm}
\label{tab:abl_sepre}
\end{table}

\begin{table}
\footnotesize
\vspace{-2mm}
\caption{Ablation of the input--output configuration on WHAMR! (SI-SNRi, dB) with the encoder-only model ($B_E$, $B_D$) = (6,0). Params. and MACs are for the single-channel models.}
\renewcommand{\tabcolsep}{2.6pt}
\def\arraystretch{1.05}
\centering
\footnotesize
\begin{tabular}{llcccccc}
\toprule
\textbf{Input} & \textbf{Output} & $(L, I)$ & $\beta$ & \textbf{Params.} (M) & \textbf{MACs} (G/s) & \textbf{1ch} & \textbf{2ch} \\
\midrule
\rowcolor{Gray}
Corr. & Filtering & (3,1) & 0.5 & 13.8 & 217.1 & \bf 18.3 & \bf 20.2 \\
Corr. & Filtering & (3,1) & 0 & 13.8 & 217.1 & 18.2 & \bf 20.2 \\
Corr. & Filtering & (3,1) & 1 & 13.8 & 217.1 & 18.2 & 20.1 \\
\midrule
Corr. & Filtering & (1,1) & 0.5 & 13.7 & 215.2 & 18.1 & 20.1 \\
Corr. & Filtering & (3,0) & 0.5 & 13.7 & 214.9 & 18.2 & 20.1 \\
Corr. & Filtering & (1,0) & 0.5 & 13.6 & 214.2 & 18.1 & 20.1 \\
Corr. & Filtering & (0,0) & 0.5 & 13.6 & 213.9 & - & 20.0 \\
\midrule
Raw & Masking & (0,0) & - & 13.6 & 213.9 & 17.9 & 19.6 \\
Raw & Filtering & (3,1) & - & 13.6 & 214.2 & 17.9 & 19.6 \\
Raw & Mapping & - & - & 13.6 & 213.9 & 17.8 & 19.4 \\
\bottomrule
\end{tabular}
\vspace{-2mm}
\label{tab:abl_corr}
\end{table}
\begin{table*}
\small
\caption{WER(\%)$\downarrow$ results for utterance-wise and continuous input evaluation on the LibriCSS. SC and LOC denote speaker counting~\cite{wang_count_2021} and localization for stream merging~\cite{taherian_leveraging_2024}, which is also used for the stitching in SR-CorrNet (Section~\ref{sec:setup_css}). Results of the compared systems are taken from the literature, except for those marked with $\ast$, which were retrained on our simulated training data.}
\renewcommand{\tabcolsep}{6.5pt}
\def\arraystretch{0.95}
\begin{center}
\footnotesize
\scalebox{0.98}{\begin{tabular}{llcccccc|cccccc}
\toprule
\multirow{2}{*}{\textbf{Method}}& \multirow{2}{*}{\textbf{Stage configuration}} &\multicolumn{6}{c|}{\textbf{Utterance-wise}}&\multicolumn{6}{c}{\textbf{Continuous}} \\
&  &{\hspace{-0.1mm}0S\hspace{-0.7mm}} & {0L} & {10} & {20} & {30} & {40} &{\hspace{-0.1mm}0S\hspace{-0.7mm}} & {0L} & {10} & {20} & {30} & {40} \\
\midrule
\multicolumn{1}{l}{No Processing}&-&11.8&11.7&18.8&27.2&35.6&43.3&15.4&11.5&21.7&27.0&34.3&40.5\\
\multicolumn{1}{l}{Oracle Sound}&-&4.9&5.1&-&-&-&-&-&-&-&-&-&-\\
\midrule
\multicolumn{14}{c}{\textit{Single-channel Evaluation}}\\
\multicolumn{1}{l}{BLSTM~\cite{chen_continuous_2020}}&SISO&12.7&12.1&17.6&23.2&30.5&35.6&17.6&16.3&20.9&26.1&32.6&36.1\\
\multicolumn{1}{l}{Conformer~\cite{chen_continuous_2021}}&SISO&12.9&12.2&15.1&20.1&24.3&27.6&13.3&11.7&16.3&20.7&25.6&29.3\\
\midrule
\multicolumn{1}{l}{U-Net~\cite{wang_multi-microphone_2021}}&SISO + SC & 9.2 & 8.9 & 11.6 & 15.5 & 20.0 & 23.1 & 9.4 & 9.7 & 11.6 & 15.2 & 19.7 & 23.0\\
\multicolumn{1}{l}{U-Net~\cite{wang_multi-microphone_2021}}&SISO-SISO + SC & 9.1 & 8.6 & 10.6 & 13.9 & 17.1 & 19.8 & 10.7&10.4&11.7& 14.8& 18.8 & 20.8\\ 
\midrule
\rowcolor{Gray}
\multicolumn{1}{l}{{SR-CorrNet-\textit{fix\textsubscript{2}}}}&SISO&{6.2}&{6.2}&{6.3}&{6.8}&{7.5}& {8.0}& \bf 7.1 & \bf 6.8 & 7.0 & 7.1 & 8.3 & 9.2 \\
\rowcolor{Gray}
\multicolumn{1}{l}{{SR-CorrNet-\textit{var\textsubscript{0-2}}}}&SISO&\bf 6.0&\bf 5.8&\bf 6.1&\bf 6.6&\bf 7.3&\bf 7.7& 7.2&\bf 6.8&\bf 6.9&\bf 6.9&\bf 8.2&\bf 9.0\\

\midrule
\multicolumn{14}{c}{\textit{Seven-channel Evaluation}}\\

\multicolumn{1}{l}{BLSTM~\cite{chen_continuous_2020}}&MISO-BF&8.3&8.4&11.6&16.0&18.4&21.6&11.9&9.7&13.4&15.1&19.7&22.0\\
\multicolumn{1}{l}{Conformer~\cite{chen_continuous_2021}}&MISO-BF&7.2&7.5&9.6&11.3&13.7&15.1&11.0&8.7&12.6&13.5&17.6&19.6\\

\multicolumn{1}{l}{U-Net~\cite{wang_multi-microphone_2021}}&MISO + SC&7.7&7.5&7.9&9.6&11.3&13.0&7.9&8.5&8.5&10.5&12.3&14.3\\
\multicolumn{1}{l}{TF-GridNet~\cite{taherian_leveraging_2024}}&MIMO&7.5&7.4&7.3&8.3&9.6&10.3&9.2&12.2&9.9&10.1&11.9&12.2\\
\multicolumn{1}{l}{TF-GridNet~\cite{taherian_leveraging_2024}}&MIMO + LOC&5.8&6.4&6.7&7.9&9.5&10.3&8.4&9.1&9.4&10.0&11.3&12.0\\
\multicolumn{1}{l}{TF-CorrNet~\cite{shin_tf-corrnet_2025}}&MISO&5.8&{5.7}&6.4&{7.3}&8.6&{9.5}&7.6&8.1&8.3&9.1&10.1&10.9\\
\multicolumn{1}{l}{TF-CorrNet~\cite{shin_tf-corrnet_2025}}&MIMO + LOC&{5.6}&{5.8}&{6.3}&{7.4}&{8.6}&{9.7}  & {6.9}&{6.4}&{7.4}&{7.8}&{9.9}&{10.8}\\
\midrule
\multicolumn{1}{l}{U-Net~\cite{wang_multi-microphone_2021}}&MISO-BF-MISO + SC&5.8&5.8&{5.9}&{6.5}&7.7&8.3&7.7&7.5&7.4&8.4&9.7&11.3\\
\multicolumn{1}{l}{TF-GridNet~\cite{taherian_leveraging_2024}}&MISO-BF-MISO&6.1&6.3&{5.9}&{6.1}&{6.7}&7.8&8.0&8.4&7.4&7.1&9.0&9.3\\
\multicolumn{1}{l}{TF-GridNet~\cite{taherian_leveraging_2024}}&MIMO-BF-MISO + LOC&{5.3}&5.7&{5.5}&{5.8}&{6.8}&7.1&6.8&6.8&{6.7}&6.9&8.4&9.0\\

\multicolumn{1}{l}{{TF-CorrNet~\cite{shin_tf-corrnet_2025}}}& MIMO-BF-MISO + LOC&{5.3}&{5.5}&{5.5}&{5.7}&{6.4}&{6.7}&  {6.4}& {6.1}& {6.2}&{6.2}&{7.4}&{7.7}\\
\midrule
\multicolumn{1}{l}{TF-GridNet$^\ast$}&MISO&6.2&5.9&6.5&7.5&8.4&9.1&7.6&6.9&7.7&8.1&9.4&10.1\\
\multicolumn{1}{l}{TF-CorrNet$^\ast$}&MISO&5.9&6.2&6.1&7.5&8.4&9.5&7.3&6.9&7.4&7.9&9.5&10.2\\
\rowcolor{Gray}
\multicolumn{1}{l}{{SR-CorrNet-\textit{fix\textsubscript{2}}}}&MISO&{5.4}&{5.6}&{5.5}&{\bf 5.4}&{6.2}& {6.5}& 6.4 & 6.5 & 6.3 & \bf 5.9 & 7.3 & 7.6 \\
\rowcolor{Gray}
\multicolumn{1}{l}{{SR-CorrNet-\textit{var\textsubscript{0-3}}}}&MISO&5.3&\bf 5.4&\bf 5.4&5.7&6.0&\bf 6.0& \bf 6.1 & \bf 5.9 & \bf 6.2 & 6.0 & \bf 6.8 & \bf 6.7\\
\rowcolor{Gray}
\multicolumn{1}{l}{\ \ + direction stitching}&MISO + LOC&\bf 5.2&\bf 5.4&5.5&5.5&\bf 5.9&\bf 6.0&6.6&6.2&6.4&\bf 5.9&7.0&6.9\\

\bottomrule
\end{tabular}}
\end{center}
\label{tab:LibriCSS}
\vspace{-2mm}
\end{table*}

\subsection{Ablation study}
\label{sec:abl}
For the WHAMR! ablations in Tables~\ref{tab:abl_sepre} and~\ref{tab:abl_corr}, the new variants were trained with the same configuration as the corresponding baselines.

\paragraph{Effects of SepRe framework}
From Table~\ref{tab:abl_sepre}, SepRe improved the separation quality over the encoder-only configuration with the same total depth, and a deeper decoder than encoder ($B_E=2$, $B_D=4$) performed better than the symmetric case ($B_E=B_D=3$), highlighting the importance of the reconstruction stage.
Even without the early supervision ($\alpha=0$), SepRe outperformed the encoder-only configuration; the early supervision, at no inference cost, and the speaker interaction module each improved it further.

\paragraph{Effects of correlation inputs for filter estimation}
In Table~\ref{tab:abl_corr}, the correlation--filtering configuration with $(L,I)=(3,1)$ achieved the highest scores under both monaural and stereo conditions, while the differences among the context lengths were modest.
The temporal and spectral extension mainly lets us use the same correlation-to-filter formulation for single-channel input. With the same filter output and receptive field, the correlation input outperformed the raw input even in the monaural condition. Filtering the raw input gave no gain over masking.
The spatial correlations were more beneficial: the gains over the raw input increased in the stereo condition, even with $(L,I)=(0,0)$.
With RMS normalization, $\beta$ had little effect: $\beta=0.5$ performed comparably to or slightly better than raw correlation ($\beta=0$) and phase-only PHAT ($\beta=1$).

\subsection{Continuous speech separation for real-recorded mixture}
\label{sec:css_result}

Table~\ref{tab:LibriCSS} compares the proposed SR-CorrNet with existing methods under both single-channel and seven-channel conditions on the LibriCSS dataset.
The systems are categorized into (i) single-stage models based on SISO, MISO, or MIMO configurations, and (ii) two-stage models (MISO/MIMO--BF--MISO) that combine beamforming and post-enhancement networks.
The proposed SR-CorrNet represents a single-stage model that internally integrates the two-stage optimization through the SepRe framework. Since the published systems were trained on different data, we additionally retrained TF-GridNet and TF-CorrNet on our simulated training data for a direct comparison (marked with $\ast$).

\paragraph{Single-channel evaluation}
A key distinction from our previous TF-CorrNet, which was designed exclusively for multi-channel input, is that SR-CorrNet naturally extends the correlation-based formulation to the single-channel case by exploiting temporal and spectral correlations in the absence of inter-channel spatial cues.
This generalization is enabled by the spatio-spectro-temporal correlation framework introduced in Section~II, where the spatial dimension reduces to a scalar when $M=1$, while the temporal and spectral context windows determined by $(L, I)$ continue to provide structured input features for filter estimation.
Whereas IF-CorrNet~\cite{shin_deep_2026} exploits inter-frame correlations for single-speaker dereverberation, SR-CorrNet jointly exploits temporal and cross-band spectral correlations for multi-speaker separation within the SepRe architecture.

Under this single-channel configuration, SR-CorrNet achieved a substantial improvement over all existing single-channel baselines.
Previous SISO systems such as Conformer and U-Net with SC exhibited considerable degradation as the overlap ratio increases, reflecting the inherent difficulty of monaural separation for real-recorded mixtures.
In contrast, SR-CorrNet maintained consistently low WERs across all overlap conditions in both utterance-wise and continuous settings, demonstrating the robustness of the proposed correlation-to-filter formulation even without spatial information.
Notably, the single-channel SR-CorrNet already surpassed or closely matched several seven-channel single-stage systems, including TF-GridNet with MIMO + LOC and TF-CorrNet with MISO configurations.
This highlights the effectiveness of the SepRe architecture combined with correlation-based filtering, which enables competitive separation quality through a simplified SISO configuration without relying on any external techniques such as beamforming, external speaker counting, or localization-based stream merging.

\paragraph{Seven-channel evaluation}
Among single-stage seven-channel methods, conventional approaches often require auxiliary modules---SC for MISO-based systems and localization (LOC) for MIMO-based systems---to stabilize separation behavior during non-overlapping or weakly overlapping segments.
Among them, TF-CorrNet with MIMO + LOC achieved the best performance in this group, confirming the advantage of correlation-based input representation for multi-channel separation.

The second group adopts explicit two-stage processing, where a beamforming stage performs coarse spatial filtering and a subsequent MISO network refines the separated streams.
This led to further improvement, particularly in the MIMO--BF--MISO configuration based on TF-CorrNet, which achieved the lowest WER among conventional frameworks.
However, the added beamforming and post-processing stages inevitably increase computational cost and system complexity.

In contrast, the proposed SR-CorrNet achieved the best results across nearly all evaluation conditions while employing a simplified single-stage MISO configuration. Under identical training data, SR-CorrNet also outperformed the retrained TF-GridNet and TF-CorrNet in all conditions, confirming that its advantage over these baselines stems from the proposed framework rather than from the training data. On average, the WERs of the dynamic split were lower than theirs by 1.6 to 2.0 points, far beyond the 95\% confidence intervals of the differences (within $\pm$0.4 points) obtained by a session-level bootstrap.
By structurally absorbing the role of post-enhancement within its internal decoder, SR-CorrNet eliminates the need for external beamforming and post-enhancement stages while maintaining superior recognition accuracy.

\begin{table}
\footnotesize
\vspace{-2mm}
\caption{Real-time factor (RTF) and MACs. We averaged RTFs over three runs on the same 3,000 seven-channel LibriCSS chunks (first channel for SISO models) using FP32 inference with batch size one on a GeForce RTX 4090. We excluded model loading, audio I/O, and warm-up from total processing time for the 2.4-s chunks and divided that time by total new-audio duration (0.8~s per chunk). We averaged MACs for the dynamic split over LibriCSS. $^\ast$Retrained as in Table~\ref{tab:LibriCSS} with the same STFT as SR-CorrNet: TF-GridNet with six blocks, an embedding dimension of 48, and 192 LSTM units, and TF-CorrNet as configured in~\cite{shin_tf-corrnet_2025}.}
\renewcommand{\tabcolsep}{5pt}
\def\arraystretch{1.05}
\centering
\footnotesize
\begin{tabular}{lccc}
\toprule
\multirow{1}{*}{\textbf{System}} & \textbf{\#Params.} (M) & \textbf{MACs} (G/s) & \multicolumn{1}{c}{\textbf{RTF}} \\
\midrule
MIMO (TF-GridNet) & 5.6 & 171.8 & 0.921\\
MIMO (TF-CorrNet) & 5.1 & 44.5 & \bf 0.070 \\
MISO (TF-GridNet$^\ast$) & 8.4 & 123.0 & 0.559\\
MISO (TF-CorrNet$^\ast$) & \bf 5.0 & \bf 17.8 & 0.080\\
\midrule
MIMO-BF-MISO (TF-GridNet) & 11.2 & 512.2 & 2.771\\
MIMO-BF-MISO (TF-CorrNet) & 10.3 & 131.8 & 0.197\\
\midrule
\rowcolor{Gray}
SISO (SR-CorrNet-\textit{fix}\textsubscript{2}) & 6.8 & 159.8 & 0.121\\
\rowcolor{Gray}
SISO (SR-CorrNet-\textit{var}\textsubscript{0-2}) & 7.4 & 119.6 & 0.099\\
\rowcolor{Gray}
MISO (SR-CorrNet-\textit{fix}\textsubscript{2}) & 7.4 & 170.4 & 0.123\\
\rowcolor{Gray}
MISO (SR-CorrNet-\textit{var}\textsubscript{0-3}) & 8.0 & 129.2 & 0.101\\
\bottomrule
\end{tabular}
\vspace{-2mm}

\label{tab:RTF_CSS}
\end{table}

\paragraph{Effect of dynamic split}
Comparing SR-CorrNet-\textit{fix\textsubscript{2}} and SR-CorrNet-\textit{var\textsubscript{0-$K_0$}}, the dynamic split module lowered WERs in 20 of the 24 conditions across the single-channel and seven-channel settings. For the seven-channel models, average WER fell from 5.8\% to 5.6\% in utterance-wise evaluation and from 6.7\% to 6.3\% in continuous evaluation. The same bootstrap gave 95\% confidence intervals of about $\pm$0.1 points for the differences. WER also improved on AMI with all recognizers (Section~\ref{sec:ami}).
The benefit was most evident in the seven-channel continuous evaluation.
In this setting, the fixed split tended to produce artificial secondary streams during single-speaker regions, causing spectral leakage that can accumulate over long-form processing.
The dynamic split mitigated this by adaptively determining the number of active speakers within each chunk, which can lead to more stable stitching across consecutive blocks.
These results confirm that the attractor-based dynamic split provides a direct and effective mechanism for handling varying speaker configurations within the separator without an external speaker counting network.
Direction-based stitching (Section~\ref{sec:setup_css}) was clearly more effective than similarity-based stitching on spontaneous AMI meetings. We also applied it to LibriCSS with a matching angle of 10$^\circ$. The methods gave comparable WERs because the static speakers' streams were already consistent enough for similarity-based stitching.

\begin{figure}
\centering
\includegraphics[width=\columnwidth]{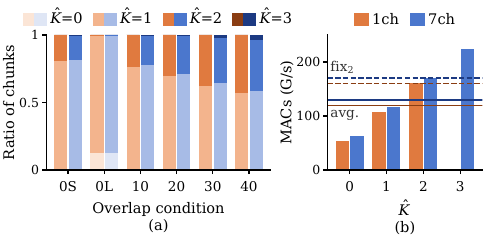}
\caption{(a) Distribution of the estimated number of speakers $\hat{K}$ for each overlap condition and (b) MACs for each $\hat{K}$ of the single- and seven-channel SR-CorrNet with the dynamic split on LibriCSS. The dashed and solid lines denote the MACs of the fixed split and the average MACs of the dynamic split, respectively.}
\label{fig:dyn_analysis}
\vspace{-2mm}
\end{figure}

\begin{table}
\footnotesize
\caption{Distribution of LibriCSS speakers by speech duration within a chunk and percentage counted by the dynamic split in each duration range.}
\label{tab:count_dur}
\renewcommand{\tabcolsep}{3.4pt}
\def\arraystretch{1.2}
\centering
\begin{tabular}{llcccccc}
\toprule
\multicolumn{2}{l}{\textbf{Duration (s)}} & $<$0.2 & 0.2--0.3 & 0.3--0.4 & 0.4--0.5 & 0.5--0.8 & $\geq$0.8 \\
\multicolumn{2}{l}{\textbf{Proportion (\%)}} & 3.5 & 1.9 & 2.0 & 1.8 & 5.7 & 85.1 \\
\midrule
\multirow{2}{*}{\textbf{Counted (\%)}} & 1ch & 8.3 & 38.1 & 57.8 & 75.1 & 89.3 & 99.6 \\
 & 7ch & 9.1 & 36.9 & 59.5 & 73.7 & 88.8 & 99.6 \\
\bottomrule
\end{tabular}
\end{table}

\paragraph{Computational efficiency}
As shown in Table~\ref{tab:RTF_CSS}, SR-CorrNet also provides a favorable quality--cost trade-off for chunk-wise processing.
While achieving comparable or superior WERs to the two-stage MIMO--BF--MISO pipelines, SR-CorrNet-\textit{fix}\textsubscript{2} operates at lower RTFs than those based on TF-CorrNet and TF-GridNet, owing to its single-stage MISO design.
At similar MACs, SR-CorrNet had substantially lower WERs and a much lower RTF than single-stage TF-GridNet. Single-stage TF-CorrNet used less computation, but its WERs were clearly higher, particularly in highly overlapped and continuous conditions.
The dynamic split further reduces average computation (Section~\ref{sec:dyn_analysis}). In all conditions, SR-CorrNet with this split matched or outperformed the two-stage TF-CorrNet pipeline in WER at similar MACs and about half its RTF.
This demonstrates the practicality of SR-CorrNet for chunk-wise continuous speech separation, achieving the effectiveness of two-stage processing within a single architecture.

\subsection{Analysis of dynamic split}
\label{sec:dyn_analysis}
\leavevmode%
On LibriCSS, the estimated number of speakers $\hat{K}$ per chunk increased with the overlap ratio (Fig.~\ref{fig:dyn_analysis}(a)). Table~\ref{tab:count_dur} shows how often a speaker was counted by speech duration within a chunk. We measured duration on the source utterance and treated the $\hat{K}$ speakers with the longest speech in each chunk as counted. The speakers with at least 0.8~s of speech, which account for most of the cases, were almost always counted, whereas those with less than 0.2~s were rarely counted. In between, most of the uncounted speakers spoke only in the history or future context, i.e., outside the output segment.
As shown in Fig.~\ref{fig:dyn_analysis}(b), the decoder is skipped for $\hat{K}=0$ and processes only $\hat{K}$ streams otherwise. On LibriCSS, the seven-channel dynamic split used 24\% less computation on average than the fixed split and had a lower RTF (Table~\ref{tab:RTF_CSS}). Average computation fell by 25\% for the single-channel model and 15--17\% for the AMI models.

\begin{table}
\footnotesize
\vspace{-2mm}
\caption{Utterance-wise WER (\%) on LibriCSS (7ch) under degradation of the reference or a non-reference channel, averaged over the six overlap conditions.}
\renewcommand{\tabcolsep}{1.8pt}
\def\arraystretch{1.05}
\centering
\footnotesize
\scalebox{0.95}{\begin{tabular}{@{}llcccccc@{}}
\toprule
\multirow{2}{*}[-2pt]{\textbf{System}} & \multirow{2}{*}[-2pt]{\textbf{Correlation}} & \multirow{2}{*}[-2pt]{\textbf{Clean}} & \multicolumn{4}{c}{\textbf{Reference channel}} & \textbf{Non-ref.} \\
\cmidrule(lr){4-7}\cmidrule(lr){8-8}
 & & & \textbf{Occl.} & \textbf{Noise} & \textbf{Phase} & \textbf{Dead} & \textbf{Dead} \\
\midrule
\multirow{3}{*}{SR-CorrNet-\textit{fix}\textsubscript{2}} & Single-ref. (\ref{eq:corr_MISO}) & 5.8 & 6.8 & 10.6 & 7.3 & 99.6 & 6.3 \\
 & Multi-ref. (\ref{eq:corr_multiref}) & 5.8 & 6.7 & 8.3 & 6.3 & 9.9 & 6.2 \\
 & \ \ + ref. dropout & 5.6 & 6.0 & 8.5 & 5.9 & 6.4 & 5.9 \\
\midrule
\multirow{3}{*}{SR-CorrNet-\textit{var}\textsubscript{0-3}} & Single-ref. (\ref{eq:corr_MISO}) & 5.6 & 6.1 & 8.7 & 6.4 & 71.8 & 5.9 \\
 & Multi-ref. (\ref{eq:corr_multiref}) & 5.6 & 6.4 & 9.0 & 6.1 & 11.6 & 6.2 \\
 & \ \ + ref. dropout & 5.7 & 6.1 & 8.7 & 5.9 & 6.4 & 6.1 \\
\bottomrule
\end{tabular}}
\vspace{-2mm}
\label{tab:ref_robust}
\end{table}

\subsection{Robustness to reference-channel degradation}
\label{sec:ref_robust}
To evaluate the sensitivity to the reference channel on real recordings, we trained the seven-channel SR-CorrNet-\textit{fix}\textsubscript{2} and SR-CorrNet-\textit{var}\textsubscript{0-3} with the multi-reference correlation in (\ref{eq:corr_multiref}), with and without a reference dropout that randomly removes microphones from the correlation computation during training. At inference, only the reference channel of LibriCSS was degraded by attenuation of 20~dB with a low-pass filter at 1~kHz (occlusion), additive white noise at 0~dB SNR (local noise), random phase rotation for each frequency bin (phase distortion), or zeroing (dead microphone), and one of the non-reference channels was also zeroed for comparison.
As shown in Table~\ref{tab:ref_robust}, all the models performed comparably for the clean reference channel. The single-reference models performed worse when the reference channel was degraded and failed for the dead microphone because every correlation feature in (\ref{eq:corr_MISO}) uses the reference channel. In contrast, the multi-reference correlation resolved the failure for the dead microphone, where the WER became close to that of the clean condition with the reference dropout, and also reduced the WERs under the other degradations in most cases. The degradation of a non-reference channel only marginally affected all the models.

\begin{table}
\footnotesize
\caption{ORC-WER (\%) on AMI with different recognizers. CSS denotes continuous processing of the whole meeting and Utt. denotes separate processing of each oracle speech segment. Sim. and Dir. denote similarity- and direction-based stitching.}
\renewcommand{\tabcolsep}{1.9pt}
\def\arraystretch{1.05}
\centering
\footnotesize
\begin{tabular}{@{}lccccccccc@{}}
\toprule
\multirow{2}{*}[-2.5pt]{\textbf{System}} & \multirow{2}{*}[-2.5pt]{\textbf{Proc.}} & \multirow{2}{*}[-2.5pt]{\textbf{Stitch.}} & \multirow{2}{*}[-2.5pt]{\textbf{Oracle}} & \multicolumn{2}{c}{\textbf{Parakeet}} & \multicolumn{2}{c}{\textbf{Whisper}} & \multicolumn{2}{c}{\textbf{Canary}} \\
\cmidrule(lr){5-6}\cmidrule(lr){7-8}\cmidrule(l){9-10}
 & & & & \textbf{Dev} & \textbf{Eval} & \textbf{Dev} & \textbf{Eval} & \textbf{Dev} & \textbf{Eval} \\
\midrule
No Processing & - & - & - & 27.2 & 27.6 & 29.8 & 32.0 & 29.2 & 31.2 \\
Headset microphone & Utt. & - & \ding{51} & 13.8 & 13.0 & 15.6 & 15.8 & 15.7 & 16.0 \\
\midrule
Beamforming~\cite{anguera_acoustic_2007} & - & - & - & 25.6 & 26.0 & 27.4 & 28.7 & 28.1 & 29.2 \\
TF-GridNet~\cite{wang_tf-gridnet_2023} & CSS & Sim. & - & 22.6 & 24.3 & 24.7 & 27.6 & 25.5 & 27.6 \\
TF-CorrNet~\cite{shin_tf-corrnet_2025} & CSS & Sim. & - & 22.8 & 24.9 & 25.2 & 28.8 & 25.7 & 28.6 \\
\rowcolor{Gray}
SR-CorrNet-\textit{fix}\textsubscript{2} & CSS & Sim. & - & 20.0 & 20.9 & 23.4 & 25.0 & 23.6 & 25.1 \\
\rowcolor{Gray}
SR-CorrNet-\textit{fix}\textsubscript{2} & CSS & Dir. & - & 17.7 & 18.6 & 21.4 & 23.0 & 21.2 & 23.0 \\
\rowcolor{Gray}
SR-CorrNet-\textit{var}\textsubscript{0-3} & CSS & Dir. & - & \bf 16.7 & \bf 17.4 & \bf 20.8 & \bf 21.9 & \bf 20.5 & \bf 21.9 \\
\midrule
GSS~\cite{boeddeker_front-end_2018} & Utt. & - & \ding{51} & 16.7 & 16.9 & 18.6 & \bf 19.7 & 19.6 & \bf 19.8 \\
\rowcolor{Gray}
SR-CorrNet-\textit{var}\textsubscript{0-3} & Utt. & Dir. & \ding{51} & 15.3 & 16.0 & \bf 18.3 & 19.9 & 19.6 & 20.7 \\
\rowcolor{Gray}
SR-CorrNet-\textit{var}\textsubscript{0-3} & CSS & Dir. & \ding{51} & \bf 15.0 & \bf 15.9 & 19.0 & 20.1 & \bf 19.2 & 20.5 \\
\bottomrule
\end{tabular}
\label{tab:ami}
\vspace{-2mm}
\end{table}

\subsection{Evaluation on AMI}
\label{sec:ami}
To further examine spontaneous conversational speech beyond the replayed recordings of LibriCSS, we evaluated the eight-channel models on AMI as described in Section~\ref{sec:setup_css}, together with the weighted delay-and-sum beamforming of BeamformIt~\cite{anguera_acoustic_2007}, TF-GridNet and TF-CorrNet trained with the same simulated room impulse responses, and guided source separation (GSS)~\cite{boeddeker_front-end_2018} with the oracle speaker activities. We also applied the direction-based stitching in Section~\ref{sec:setup_css} with a matching angle of 30$^\circ$, where a direction is kept by only one stream so that each speaker is tracked by a single stream. For simulated mixtures of two or three speakers at different positions, the direction head of SR-CorrNet-\textit{var}\textsubscript{0-3} estimated 88.1\% of the active frames within 5$^\circ$ of the true directions.

In Table~\ref{tab:ami}, with Parakeet-TDT, SR-CorrNet outperformed the beamforming as well as TF-GridNet and TF-CorrNet with the same similarity-based stitching, and the direction-based stitching and the dynamic split successively reduced the ORC-WERs on both the development and evaluation sets. SR-CorrNet, trained only on simulated mixtures, used no diarization or oracle speaker activity information. Its performance approached that of GSS using oracle speaker activities.

For the comparison in the last block of Table~\ref{tab:ami}, we used the same oracle speaker activities as GSS to set the number of SR-CorrNet output streams for each chunk. We processed each reference speech segment separately as in GSS or processed the whole meeting continuously. With Parakeet-TDT, SR-CorrNet yielded clearly lower ORC-WERs than GSS with both the segment-wise and the continuous processing.

With all recognizers (Table~\ref{tab:ami}), SR-CorrNet outperformed beamforming, TF-GridNet, and TF-CorrNet. Direction-based stitching and the dynamic split reduced ORC-WERs on both sets with each recognizer. The oracle number of speakers also reduced ORC-WERs in each case, giving results comparable to GSS with Whisper and Canary. Headset microphone ORC-WERs for the same oracle segments give a reference upper bound (Table~\ref{tab:ami}, second row).

\section{Conclusion}
We presented SR-CorrNet, a unified TF-domain separation framework that combines the SepRe strategy with the spatio-spectro-temporal correlation-to-filter formulation, which generalizes our previous TF-CorrNet to single-channel inputs and to multiple reference channels for robustness against reference-channel degradation.
An attractor-based dynamic split module further integrates speaker counting and conditioning into the separator without an external counting network while reducing the average computation.
Experiments on WSJ0-\{2,3,4,5\}mix, WHAMR!, LibriCSS, and AMI verified the effectiveness of the proposed approach across anechoic, noisy-reverberant, and real-recorded conditions.
On LibriCSS, SR-CorrNet achieved the lowest WERs in nearly all conditions with a simplified single-stage configuration, also outperforming the baselines retrained on identical data.
On AMI, SR-CorrNet trained only on simulated mixtures approached the performance of guided source separation with oracle speaker activities without any diarization information.

\textbf{In the future,} we plan to extend the correlation-based formulation to arbitrary array geometries and to incorporate long-context speaker identity modeling for joint speaker diarization and speech separation.

\ifCLASSOPTIONcaptionsoff
  \newpage
\fi

\bibliographystyle{IEEEtran}
\setstretch{1.0}
\bibliography{references.bib}

\end{document}